\documentclass[11pt]{article}
\usepackage[letterpaper, margin=1in]{geometry}
\usepackage{setspace} 
\usepackage{graphicx}
\usepackage[dvipsnames,svgnames,x11names]{xcolor}
\usepackage{cite}
\usepackage{booktabs}
\usepackage{amsmath,mathtools,amssymb}
\usepackage{textcomp,gensymb}
\usepackage{etoolbox}
\usepackage{multirow}
\usepackage[pdfusetitle]{hyperref}
\usepackage{authblk}

\renewcommand{\arraystretch}{0.5}
\makeatletter
\patchcmd{\env@cases}{1.2}{0.72}{}{}
\makeatother

\newcommand{\email}[1]{\footnote{\href{mailto:#1}{#1}}}
\hypersetup{
    colorlinks,
    linkcolor={red!50!black},
    citecolor={blue!50!black},
    urlcolor={blue!80!black}
}

\title{\Large\bf%
Rolling Down the Leptonic BSM Landscape Using Machine Learning Techniques}
\author[1,2]{Alfredo Aranda\email{fefo@ucol.mx}}
\author[3]{Raymundo Ramos\email{raramos@kias.re.kr}}
\author[1,2]{Alexander J. Stuart\email{astuart@ucol.mx}}
\affil[1]{\textit{Facultad de Ciencias, Universidad de Colima, C.P. 28045, Colima, México}}
\affil[2]{\textit{Dual CP Institute of High Energy Physics, C.P. 28045, Colima, México}}
\affil[3]{\textit{Quantum Universe Center, Korea Institute for Advanced Study, Seoul 02455, Korea}}

\date{}

\begin{document}

\maketitle

\begin{abstract}
    \noindent%
    In this work, we adapt and apply techniques from machine learning
    to the exploration of physics beyond the Standard Model in the leptonic sector.
    Namely, we employ initialization and optimization,
    as they are applied in machine learning,
    to minimize a loss function
    that describes textures or conditions
    which we want in the neutrino mass matrix.
    The model free parameters are explored during the optimization, and after training for a number of optimization steps, we obtain matrices that approximately follow the desired forms,
    as well as their corresponding optimized parameters.
    We also discuss extensions and additional applications of the ideas presented here
    in conjunction with other methods based on artificial intelligence.
\end{abstract}

\newpage

\section{Introduction}
\label{sec:intro}

The Standard Model (SM), inspired by naturalness, simplicity, and experimental results,
has endured precise observations as performed up to these
days~\cite{ParticleDataGroup:2024cfk}.
While this indicates that all the elements of the SM are needed
to explain these observations,
other phenomena point towards components of nature that remain unexplained.
One notable unexplained component is the Yukawa sector,
where masses of families of quarks and leptons depend on many free parameters
that, rather than being predicted, are set to follow experimental constraints.
Moreover, in the leptonic sector,
neutrino masses are absent in the SM
while observations of neutrino oscillations
indicate that they are massive and that their
mass eigenstates do not correspond to the same flavor eigenstates
of the charged leptons~\cite{Mann:1976mp,Wolfenstein:1978uw}.
Having massive neutrinos that oscillate
complicates any formulation beyond the SM (BSM) that simply adds fields to generate a few additional mass terms in the SM Lagrangian.
Instead, we have to deal with diagonalizations of complex matrices,
new particles with their own masses and associated energy scales,
breaking of new symmetries,
and many more characteristics.
It is clear that attempting to include masses for neutrinos in (BSM)
theories will unavoidably add new parameters, as they must follow
constraints from the aforementioned observations.

Interestingly, perceptive observations of experimental results
may give insights into the characteristics of an underlying behavior, e.g.,
a nearly diagonal (quark) mixing matrix
may indicate high-energy symmetries that kept
particles (quarks) from mixing until it was broken at a lower energy scale.
The leptonic sector is not an exception,
with ``bare'' mixing normally identified with a form known as \emph{tribimaximal mixing}~\cite{Harrison:2002er,Harrison:2002kp,Xing:2002sw,He:2003rm,Xing:2006xa}.
The number of BSM proposals
that attempt to repeat this form of the mixing matrix
via additional particles and/or symmetries keeps growing
even after the reactor mixing angle, $\theta_{13}$, has been measured as non-vanishing~\cite{DayaBay:2012fng,RENO:2012mkc,DoubleChooz:2014kuw}.
These proposals have been made
by a mixture of human experience in combination with trial and error.
In this work, we want to explore the possibility of using new computational techniques
from the artificial intelligence (AI) era to automatize the discovery of hints
of properties of an underlying model from known experimental results.
In particular, our interest lies in the direction of
opening doors to comprehensive and robust
analysis of a number of possible configurations that may be discovered,
including statistical analysis on their frequency and determination of their predictive power.
One could imagine the possibility of starting with a far too general model
and using some form of optimization to infer relationships among parameters,
potentially pointing towards simpler formulations.
Our approach will follow the skeleton devised in
Refs.~\cite{Matchev:2023mii} and~\cite{Matchev:2024ash},
although we will develop our own loss functions and training
system.\footnote{
    For other applications of machine learning methods to leptons, quarks, and flavor in BSM physics see Refs.~\cite{
        Nishimura:2020nre,Nishimura:2024apb,Nishimura:2025rsk,Nishimura:2025knz,Giarnetti:2025mit}.
}

To be more precise, we want to focus on a particular and important part of AI,
known as \emph{machine learning} (ML).
The process of ML itself is a mathematical one that, put in very simple terms,
aims to optimize the parameters of a mathematical function typically identified as a model,
via the minimization of a loss function.
In this work, our purpose is to use techniques
that empower ML with objects
found in theoretical setups of the leptonic sector.
Towards this purpose,
we will formulate elements from the theoretical leptonic model
as a function delivering predictions for several observables.  
The variables of the model will be employed as those optimized by typical algorithms
used in ML, and, as is done in ML,
we will formulate loss functions where the minimum is achieved
when the output of the model is consistent with our desired result.
However, an important difference in our case is that,
instead of obtaining a final reusable model that encodes inputs to outputs,
we obtain a number of neutrino mass matrices
with (approximately) the conditions contained in the loss function.
In addition, we obtain optimized values for variables of the model
that were used during the \emph{learning} process.

This paper is structured as follows.
In Section~\ref{sec:nusmbsm}, we give a brief description
of the leptonic sector of the SM
and simple extensions of it involving massive neutrinos,
as well as the details needed for the numerical setup.
In Section~\ref{sec:nulearn}, we present all the details
of the process used to optimize the variables involved
and obtain particular forms of the neutrino mass matrix.
This section is where analogies to ML methods
are developed.
In Section~\ref{sec:discussion}, we discuss interesting points
of the results obtained during the numerical process,
make some connections to theoretical details,
and comment on future improvements.
Finally, in Section~\ref{sec:conclusion},
we present the conclusions of this work.

\section{Neutrinos in the Standard Model and Beyond}
\label{sec:nusmbsm}

The neutrinos are part of the leptonic sector of the SM\@.
They are quite unique as they do not have charge
and are the only fermionic fields that do not have a right-handed field.
Due to this, and the accidental symmetries of the SM,
it is impossible to generate neutrino masses at tree or loop level
using only what is contained in the SM~\cite{ParticleDataGroup:2024cfk}.
To see this explicitly, observe that in the SM, neutrinos are part of $SU(2)$ lepton doublets, i.e., 
\begin{equation}
    \label{eq:Ldoublet}
    L_{\ell L} = \begin{pmatrix}
        \nu_{\ell L} \\
        \ell_L
    \end{pmatrix}\, , \quad \ell = e, \mu, \tau \,,
\end{equation}
with the left-handed neutrinos in the upper component
and the left-handed charged leptons in the lower component.
In the SM, masses for fermions are generated from Yukawa couplings of the $L_{\ell L}$ doublets,
with the Higgs scalar doublet, $\phi$,
and the right-handed fermion singlets, i.e., 
\begin{equation}
    \label{eq:clepyukawas}
    -\mathcal{L}_Y = \sum_{j, k} Y^\ell_{jk}\, \bar{L}_{j L}\, \phi\, e_{k R} + \text{H.c.}\,,
\end{equation}
where indices, $j$ and $k$, run over the three lepton families of Eq.~\eqref{eq:Ldoublet}.
When the doublet, $\phi = (\phi^+, \phi^0)^T$,
acquires the vacuum expectation value,
$\langle \phi \rangle = (0, v/\sqrt{2})$, charged leptons acquire the masses
\begin{equation}
    \label{eq:clepmasses}
    m^\ell_{jk} = Y^\ell_{jk} \frac{v}{\sqrt{2}}\,.
\end{equation}
Yet, in the SM, there is no right-handed neutrino field analogous to $e_{\ell R}$,
and, therefore, this mechanism cannot be used to generate masses for neutrinos. Additionally, the
global accidental symmetries of total lepton number
and $B - L$, forbid the generation of mass via the product of two $L_{\ell L}$~\cite{ParticleDataGroup:2024cfk}.

\subsection{Dirac Neutrinos}

In this work, we explore a minimalist extension of the SM by respecting its accidental symmetries and introducing three right-handed neutrinos that generate a $3\times 3$ Dirac neutrino mass matrix, $M^\nu_D$.
This mass matrix is diagonalized by two unitary matrices, $V^R$ and $W^\nu$, i.e., 
\begin{equation}
    \label{eq:mnudiag}
    V^{R\dag} M^\nu_D W^\nu = \mathrm{diag}(m_1, m_2, m_3),
\end{equation}
where $m_1$, $m_2$, and $m_3$ are the neutrino mass eigenstates.
The Pontecorvo-Maki-Nakagawa-Sakata mixing matrix ($U_\text{PMNS}$) of the leptonic sector is defined as\cite{ParticleDataGroup:2024cfk}
\begin{equation}
    U_\text{PMNS} = \mathcal{P}^\dagger_\ell U^{\ell\dagger} W^\nu = \mathcal{P}^\dagger_\ell U^{\ell\dagger} U^\nu \mathcal{P}_\nu \, ,
\end{equation}
where we have added $U^{\ell}\mathcal{P}_\ell$,
the analogous of $W^\nu$ for charged leptons,
and also expanded $W^\nu = U^\nu \mathcal{P}_\nu$.
The $\mathcal{P}_{\ell,\nu}$ matrices are diagonal phase matrices.
In the case where charged leptons
are assumed to be already in their mass eigenstates
$U^{\ell\dagger}$ becomes the identity.
Additionally,
$\mathcal{P}_\ell$ can be removed
through unphysical charged-lepton phase redefinition.
In the case of $\mathcal{P}_\nu$,
it is non-removable only for Majorana neutrinos.
Therefore, by
assuming that charged leptons are already in their diagonal basis
and redefined to absorb $\mathcal{P}_\ell$
and that neutrinos receive only Dirac masses ($W^\nu \to U^\nu$),
 $U_\text{PMNS}$ becomes
\begin{equation}
    \label{eq:Upmns}
    U_\text{PMNS} = \mathcal{P}^\dagger_\ell U^{\ell\dagger} U^\nu \mathcal{P}_\nu \to U^\nu\, .
\end{equation}
The elements of this matrix can be expressed using the Particle Data Group (PDG) parameterization
given by
\begin{equation}
\label{eq:Upmnsparam}
U_\text{PMNS} = \left(
\begin{array}{ccc}
    c_{12} c_{13} & s_{12} c_{13} & s_{13} e^{-i\delta^{CP}} \\
    -s_{12} c_{23} - c_{12} s_{13} s_{23} e^{i\delta^{CP}} &  c_{12} c_{23} - s_{12} s_{13} s_{23} e^{i\delta^{CP}} & c_{13} s_{23} \\
    s_{12} s_{23} - c_{12} s_{13} c_{23} e^{i\delta^{CP}} & -c_{12} s_{23} - s_{12} s_{13} c_{23} e^{i\delta^{CP}} & c_{13} c_{23}
\end{array}
\right) \,,
\end{equation}
where $s_{jk} = \sin\theta_{jk}$ and $c_{jk} = \cos\theta_{jk}$.
Using Eqs.~\eqref{eq:mnudiag} and~\eqref{eq:Upmns}, we obtain 
\begin{equation}
    \label{eq:mnuundiag} 
    M^\nu_D = V^{R}\mathrm{diag}(m_1, m_2, m_3) W^{\nu\dag}
         =  V^{R}\mathrm{diag}(m_1, m_2, m_3) U_\text{PMNS}^{\dag} \,.
\end{equation}
This allows us to obtain the elements of $M^\nu_D$
by setting the values of $U_\text{PMNS}$ to the measured oscillation parameters, and by writing the elements of $V^R$ using the parameterization\footnote{
    Note that the parameterization of $\mathcal{P}^R$
    is compatible with the diagonal phase matrix
    used by the Particle Data Group review up to 2018~\cite{ParticleDataGroup:2018ovx}
    and can be easily related to the parameterization used in more recent versions of the same review.
}
\begin{equation}
    \renewcommand*{\arraystretch}{0.8}
    \label{eq:Vrparams}
    V^R =
    \begin{pmatrix}
        c^R_{12} c^R_{13} & s^R_{12} c^R_{13} & s^R_{13} e^{-i\delta^R} \\
        -s^R_{12} c^R_{23} - c^R_{12} s^R_{13} s^R_{23} e^{i\delta^R} &  c^R_{12} c^R_{23} - s^R_{12} s^R_{13} s^R_{23} e^{i\delta^R} & c^R_{13} s^R_{23} \\
        s^R_{12} s^R_{23} - c^R_{12} s^R_{13} c^R_{23} e^{i\delta^R} & -c^R_{12} s^R_{23} - s^R_{12} s^R_{13} c^R_{23} e^{i\delta^R} & c^R_{13} c^R_{23}
    \end{pmatrix}
    \times \mathcal{P}^R \ ,
\end{equation}
where $s^R_{jk} = \sin\theta^R_{jk}$ and $c^R_{jk} = \cos\theta^R_{jk}$
and $\mathcal{P}^R = \mathrm{diag}(1, \exp(i \phi^R_2), \exp(i \phi^R_3))$.
Additionally, we need to fix the neutrino mass eigenvalues
which we can do by choosing the mass of the lightest eigenstate and
using squared mass differences to fix the other two masses.
For example, if we choose $m_1$ to be the lightest mass 
(normal hierarchy),
we can fix the other masses using
\begin{align}
    \label{eq:mnu2}
    m_2 & = \sqrt{\Delta m^2_{21} + m_1^2}\,,\\
    \label{eq:mnu3}
    m_3 & = \sqrt{\Delta m^2_{31} + m_1^2}\,.
\end{align}
When using these equations, we will always consider $m_1$ as real, resulting in $m_2$ and $m_3$ being real as well.
We will leave the contribution from complex masses as part of the phases in $V^R$.

At some point, we will need to obtain oscillation parameters starting from numerical $M^\nu_D$.
In that case, we will proceed by first obtaining $V^R$ and $U^\nu$ from squaring Eq.~\eqref{eq:mnudiag}
using
\begin{align}
    \label{eq:VRfromsquaremnu}
    V^{R\dagger} M^\nu_D M^{\nu\dagger}_D V^R & = \mathrm{diag}(|m_1|^2, |m_2|^2, |m_3|^2)\,, \\
    \label{eq:Unufromsquaremnu}
    U^{\nu\dagger} M^{\nu\dagger}_D M^\nu_D U^\nu & = \mathrm{diag}(|m_1|^2, |m_2|^2, |m_3|^2)\,,
\end{align}
and obtaining numerical $V^R$ and $U^\nu$
from the eigenvectors of $M^\nu_D M^{\nu\dagger}_D$ and $M^{\nu\dagger}_D M^\nu_D$,
respectively.
The values of $|m_1|^2$, $|m_2|^2$, and $|m_3|^2$ are obtained from the eigenvalues
of either $M^\nu_D M^{\nu\dagger}_D$ or $M^{\nu\dagger}_D M^\nu_D$.
We make sure that the obtained $V^R$ and $U^\nu$ result in real masses
by redefining $V^R$ such that $V^{R\dagger} M^\nu_D U^\nu$ is real.
From the numerical $U^\nu$
(that was equalized to $U^\text{PMNS}$ in Eq.~\eqref{eq:Upmns})
we can obtain mixing angles using the parameterization of Eq.~\eqref{eq:Upmnsparam}
and the following expressions
\begin{equation}
    \label{eq:Upmnsangles}
    \theta_{23} = \tan^{-1} \left( \frac{|U_{\mu 3}|}{|U_{\tau 3}|} \right)\,,\quad
    \theta_{12} = \tan^{-1} \left( \frac{|U_{e 2}|}{|U_{e 1}|} \right)\,,\quad
    \theta_{13} = \sin^{-1} \left( |U_{e 3}| \right)\,.\quad
\end{equation}
The amount of $CP$ violation can be quantified using the Jarlskog invariant~\cite{Jarlskog:1985ht,Wu:1985ea}
given by
\begin{equation}
    \label{eq:jarlskog}
    J = \mathrm{Im}\left(U^\nu_{e1} U^\nu_{\mu 2} U^{\nu\dagger}_{e 2} U^{\nu\dagger}_{\mu 1}\right)\,,
\end{equation}
which can be rewritten as
\begin{equation}
    \label{eq:Upmnsjarlskog}
    J = s_{23} c_{23} s_{12} c_{12} s_{13} c_{13}^2 \sin\delta^{CP}\,,
\end{equation}
using the PDG parameterization of Eq.~\eqref{eq:Upmnsparam}.
From Eqs.~\eqref{eq:jarlskog} and~\eqref{eq:Upmnsjarlskog}, we can obtain the $\delta^{CP}$ phase by using
the numerical values of $U^\nu$ and the angles obtained from using Eq.~\eqref{eq:Upmnsangles} as
\begin{equation}
    \label{eq:arcsinejarlskog}
    \delta^{CP} = \sin^{-1} \left[
        \frac{
            \mathrm{Im}\left(U^\nu_{e1} U^\nu_{\mu 2} U^{\nu\dagger}_{e 2} U^{\nu\dagger}_{\mu 1}\right)
        }{
            s_{23} c_{23} s_{12} c_{12} s_{13} c_{13}^2
        }
    \right]\,.
\end{equation}
Note that the range of the arcsine function is $[-\pi/2, \pi/2]$ or $[-90\degree, 90\degree]$.
Thus, sometimes we will need an extra step to obtain the correct quadrant.
This step will be explained when it becomes necessary later in this work.
Finally, the value that we need to check against KATRIN's measurement is obtained
from the eigenvalues of $M^\nu_D$ and the first row of $U^\nu$
using~\cite{KATRIN:2024cdt}
\begin{equation}
    \label{eq:mnukatrin}
    m_\nu = \sqrt{\sum_{j=1}^{3} |U^\nu_{ej}|^2 m_j^2}\,.
\end{equation}
The numerical values
determined in a recent global fit
by the NuFIT collaboration~\cite{Esteban:2024eli}
are summarized in Table~\ref{tab:measurements}
assuming normal ordering.
The 2$\sigma$ ranges have been determined by interpolating
the 1-dimensional projections provided in the same global fit.

\begin{table}[tb]
    \renewcommand{\arraystretch}{1.5}
    \setlength{\tabcolsep}{0.4cm}
    \center
    \begin{tabular}{lcccc}
        \toprule
                                                  &  bfp $\pm 1\sigma$           &  2$\sigma$ range  \\
        \cmidrule(r){1-1}                         \cmidrule(lr){2-3}
        $s^2_{12}$                                & $0.3088^{+0.0067}_{-0.0066}$ & [0.2957,0.3224] \\
        $\theta_{12}$ [$\degree$]                 & $33.76^{+0.42}_{-0.41}$      & [32.94,34.60] \\
        $s^2_{23}$                                & $0.470^{+0.017}_{-0.014}$    & [0.444,0.575] \\
        $\theta_{23}$ [$\degree$]                 & $43.27^{+1.0}_{-0.82}$       & [41.76,49.29] \\
        $s^2_{13}$                                & $0.02249 \pm 0.00057$        & [0.02133,0.02363] \\
        $\theta_{13}$ [$\degree$]                 & $8.62 \pm 0.11$              & [8.40,8.84] \\
        $\delta^{CP}$ [$\degree$]                 & $207^{+23}_{-20}$            & [145,349] \\
        $\Delta m^2_{21}$ $\left[10^{-5} \text{ eV}^2\right]$ & $7.537^{+0.094}_{-0.10}$     & [7.337,7.728] \\
        $\Delta m^2_{31}$ $\left[10^{-3} \text{ eV}^2\right]$ & $2.521^{+0.026}_{-0.018}$    & [2.479,2.569] \\
        \midrule
        $m_\nu < 0.45$\ eV & & \\
        \bottomrule
    \end{tabular}
    \caption{\label{tab:measurements}%
        Numbers used for numerical inputs and constraints.
        The numbers in the second column
        correspond to the numbers reported
        in the most recent global fit
        made by the NuFIT collaboration (version 6.1)~\cite{nufitwebsite}.
        The 2$\sigma$ range has been interpolated from the 1-dimensional projections
        provided in the same global fit.
        The value for the upper bound of $m_\nu$ was obtained by the KATRIN experiment~\cite{KATRIN:2024cdt}.
    }
\end{table}

\section{Learning Parameters of the Leptonic Sector}
\label{sec:nulearn}

In this section, we implement ML techniques
to find and enforce conditions between elements in the neutrino mass matrix.
We do this by starting from randomizing theoretical and experimental inputs.
The main idea is to begin by setting some values using experimentally measured
neutrino oscillation parameters
and continue by randomizing model parameters.
Those randomized model parameters are the variables that we will attempt to
\emph{learn}
by crafting loss functions which will be minimized.
The loss functions will contain conditions that we wish
to impose on the neutrino mass matrix,
although we will try to stay on simple
conditions that enforce some form of coincidences between elements.
We will also repeat the process several times and
pay special attention to the most common cases
or cases where we find interesting and unexpected coincidences.

The first step is the initialization of the involved variables and parameters.
This step takes care of setting the initial values of the parameters
before any optimization process is used.
In the case of the measured oscillation parameters,
we initialize them from gaussian distributions using their central value
and the 1$\sigma$ error.  In the cases where the $\pm 1\sigma$ errors are not symmetric,
we use the smallest error.
In the case of $\theta_{13}$ and $\Delta m^2_{31}$,
considering their rather small error bars,
we have fixed them to their best fit central values.
The numbers we use are based on the NuFIT 6.1 global fit~\cite{Esteban:2024eli,nufitwebsite}.
Additionally, the mass of the lightest neutrino mass eigenstate
can be constrained from above using the earth based experiment KATRIN,
which sets an upper bound $m_\nu < 0.45$~eV,
where $m_\nu^2 = \sum_j |U_{ej}|^2 m_j^2$~\cite{KATRIN:2024cdt}.
This result is independent of mass ordering
and issues with the cosmological model.
The rest of the variables are the free parameters of $V^R$,
given by the angles $\theta^R_{jk}$, with $jk=12,13,23$,
and the phases $\delta^R$, $\phi^R_2$ and $\phi^R_3$.
For these variables we use a random uniform distribution,
with the angles $\theta^R_{jk}$ being in the $[0, \pi/2]$ range,
and the phases in the $[-\pi, \pi]$ range.
Considering that the measured oscillation parameters are not completely free to vary,
we will apply optimization using only the parameters of $V^R$ and, sometimes,
the lightest neutrino mass, $m_1$.
With all these parameters given numerical values,
we can obtain a numerical value for $M^\nu_D$
using Eq.~\eqref{eq:mnuundiag}.

The second step after initialization
is defining a loss function that will be minimized during the optimization process.
This loss function should reflect the conditions that we want to enforce
during the optimization process,
with its minimum defining the result that we want to obtain.
For example, if we want to obtain textures with equal entries,
the loss function should be defined such that its minimum occurs when the selected entries
are equal.
As another example, assume we want to enforce some elements to have a particular value.  
Then, the minimum of the loss function should be at the point where those elements
have acquired exactly that particular value.
In all cases, the loss function should increase in value
as we move away from the desired conditions.
In this work, we will define the loss as a function of the matrix elements $(M^\nu_D)_{jk}$ and some case-dependent additional parameters.

When a loss function has been defined,
we can proceed to optimize the variables of our problem.
In our case, these variables are the angles $\theta^R_{jk}$,
and the phases $\delta^R$, $\phi^R_2$ and $\phi^R_3$.
Additionally, we may include $m_1$ when relevant.
There are several optimization algorithms,
but most of them work by applying small variations to the variables that we want to optimize
and calculate the change in the loss function from these small variations.
After this, the values of the variables are updated in the direction where the small variations
were found to have the steepest drop in the loss function.
This is repeated for a number of steps until the loss function can be considered minimized.
Note that there may be complications when several local minima exist
as the optimization process can get stuck in any one of them.
Additionally, if the loss function is too flat, the optimization may require
many steps to converge to the global minimum.
These problems may be somewhat alleviated by adjusting the parameters of the optimization
process or by making adjustments to the loss function.
We will comment about these problems in the cases where they are found in our specific examples.

After the optimization process has worked for a number of steps
and the loss function does not seem to yield a better minimum,
we proceed to analyze the obtained mass matrix to determine if our goal
has been reached and to analyze the obtained parameters.
Since we are working directly with the parameters of $V^R$
and not with the parameters of a very large model,
we can easily present the optimized variables and provide some insights
on the relationship between some neutrino textures
and $V^R$.

\begin{figure}[htb]
    \centering
    \includegraphics{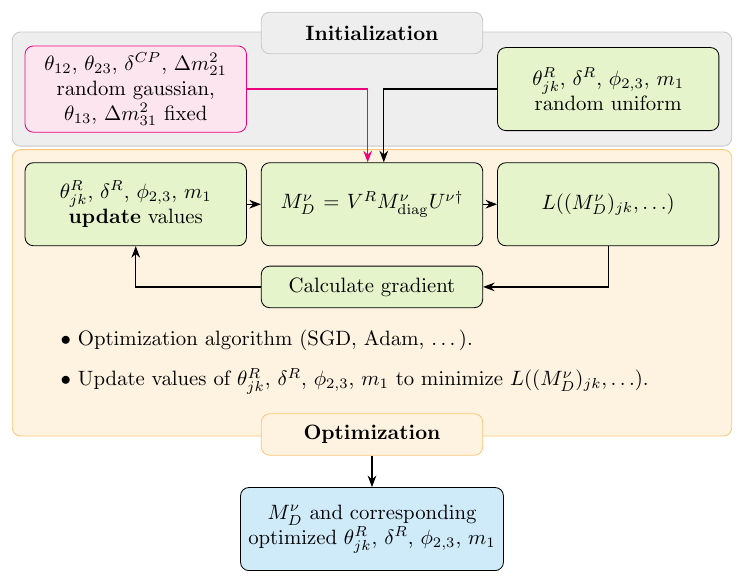}%
    \caption{\label{fig:genchart}%
        Flowchart of the optimization process
        used to obtain desirable forms of $M^\nu_D$
        and the corresponding parameters.
    }
\end{figure}

\subsection{Loss Function}

The most relevant point connecting the theory described in Sec.~\ref{sec:nusmbsm}
and the learning process described in Fig.~\ref{fig:genchart}
is the \emph{loss function}.
Typically,
in ML,
the loss function is determined by the objective of the model being trained.
For example, when training a model for binary classification, one generally
chooses the binary cross entropy loss,
while for models that have arbitrary output, one may choose
other functions such as the mean of squared or absolute errors.
In our case,
the objective is to find the parameters for the combination
$V^{R}\mathrm{diag}(m_1, m_2, m_3) U_\text{PMNS}^{\dag}$
that result in a desirable form of $M^\nu_D$.
From this idea, it is clear that the loss function must
contain some of the elements of $M^\nu_D$ in some form.
Additionally,
experimental constraints can be included in the loss function
to direct the optimization process towards viable values of the parameter space.
In our case, we are already enforcing values for the oscillation parameters
of $U_\text{PMNS}$ as part of the initialization.
However,
when including the lightest neutrino mass, $m_1$,
as an optimizable parameter,
we will need to make sure that the values of the neutrino masses
remain below the upper limit set by experimental measurements.
As previously mentioned,
we consider the constraint of the effective electron-antineutrino mass
obtained by KATRIN, i.e.,  $m_\nu < 0.45$~eV~\cite{KATRIN:2024cdt},
where $m_\nu^2 = \sum_i |U_{ei}|^2 m_i^2$.
The corresponding term in the loss function is given by
\begin{equation}
    \label{eq:losskatrin}
    L_\text{KATRIN} = w_K \max(m_\nu - 0.45\text{\ eV}, 0)^2,
\end{equation}
where $w_K$ is an adjustable weight that will be set to a large number.
If one is worried about this term being dimensionful,
the weight $w_K$ could be thought as having units of inverse squared mass.
Note that, in our case,
$L_\text{KATRIN}$ depends on measured oscillation parameters
and $m_1$ through obtaining neutrino physical masses using Eqs.~\eqref{eq:mnu2} and~\eqref{eq:mnu3}.
Additionally,
we will consider different forms of the loss function that
directly include elements of $M^{\nu}_D$.
The first loss function we consider
enforces equality of two elements and is given by
\begin{equation}
    \label{eq:loss1diff}
    L_{1\text{diff}} = \min\left(
        \frac{\left|\left(M^\nu_{D}\right)_{jk} - \left(M^\nu_{D}\right)_{lm}\right|}
            {\left|\left(M^\nu_{D}\right)_{jk}\right| + \left|\left(M^\nu_{D}\right)_{lm}\right|}
    \right)\,,
\end{equation}
where the minimum of all possible differences is taken after initialization
but before optimization.
This means that the values of the $j$, $k$, $l$ and $m$ indices
depends on the initialization of the parameters.
Note that this loss function equalizes the two mass matrix elements
down to their real and imaginary components.

Another option we consider
is the presence of vanishing elements
in the neutrino mass matrix.
Thus, we consider a simple loss function that requires us to decide on a series of vanishing elements
before optimization, i.e., 
\begin{equation}
    \label{eq:lossnzero}
    L_{n\text{-zero}} = \frac{
        (9 - n)\sum_{jk\in \text{vanish}} |\left(M^\nu_{D}\right)_{jk}|
}{
    n\sum_{lm\in\text{non-zero}} |\left(M^\nu_{D}\right)_{lm}|
}\,,
\end{equation}
where $n$ is the number of vanishing elements,
$jk$ runs over the elements that were chosen to vanish,
while $lm$ runs over the elements that are expected to be non-zero.
Note that this loss function may not always result in minimization of all the $n$ elements included
in the numerator.

We also explore
the possibility of letting the optimization process itself
decide which elements vanish.
Our approach in this case is different to the two options before
as we will not be choosing particular elements before optimization.
Rather, we design loss functions that are capable of counting the number of elements
with absolute value above and below some threshold and use optimization
to increase the number of elements below this threshold while also reducing their absolute value.
We consider the elements above the threshold as non-vanishing
while the elements below threshold are considered as possible good candidates for vanishing elements.
At this point, we have two complementary options: count the number of elements below threshold
or the number of elements above it.
If we have a mass matrix $M$, the functions that can achieve this counting
have the following asymptotic behavior:
\begin{align}
    \label{eq:fvcount}
    f_\text{vc}(M_{jk}; \epsilon_\text{vc}, \ldots) & \approx \begin{cases}
        0 & \text{if } \left|M_{jk}\right| \gg \epsilon_\text{vc}\,, \\
        1 & \text{if } \left|M_{jk}\right| \ll \epsilon_\text{vc}\,,
    \end{cases} \\
    \label{eq:fnvcount}
    f_\text{nvc}(M_{jk}; \epsilon_\text{nvc}, \ldots) & \approx \begin{cases}
        0 & \text{if } \left|M_{jk}\right| \ll \epsilon_\text{nvc}\,, \\
        1 & \text{if } \left|M_{jk}\right| \gg \epsilon_\text{nvc}\,,
    \end{cases}
\end{align}
where $\epsilon_\text{vc}$ and $\epsilon_\text{nvc}$ are thresholds mentioned earlier.
The ellipsis serves as a placeholder for additional parameters that the actual $f_\text{vc,nvc}$ may have.
These functions need to be applied to all the elements of the mass matrix
and then added to a total loss function of the form
\begin{align}
    \label{eq:lossvcount}
    L_\text{vc}(M; \epsilon_\text{vc}, \ldots) & = \left| n_\text{max} - \sum_{j,k} f_\text{vc}(M_{jk}; \epsilon_\text{vc}) \right|\,, \\
    \label{eq:lossnvcount}
    L_\text{nvc}(M; \epsilon_\text{nvc}, \ldots) & = \sum_{j,k} f_\text{nvc}(M_{jk}; \epsilon_\text{vc})\,,
\end{align}
where $n_\text{max}$ in Eq.~\eqref{eq:lossvcount} can be used to limit the number of elements.
In principle, we can also limit the number of elements above threshold in Eq.~\eqref{eq:lossnvcount},
but in practice it will hardly reach zero
as that would mean that all the elements have been brought considerably below threshold.
For a randomly initialized $3\times 3$ matrix,
the sums in Eqs.~\eqref{eq:lossvcount} and~\eqref{eq:lossnvcount}
will most likely start as 0 and 9, respectively,
unless the respective $\epsilon$ threshold is chosen already among the same size as the elements of the matrix.
In Eq.~\eqref{eq:lossvcount},
the optimization process will bring the sum closer to $n_\text{max}$,
while for Eq.~\eqref{eq:lossnvcount}
it will attempt to reduce the value of the sum as much as possible.
As usual, there is the possibility that minimizing any loss function
with these forms gets stuck in a flat region or a local minimum,
which will be consistent with textures with a particular number of vanishing elements.
This is actually advantageous,
as after many attempts it will allow the discovery of candidate textures
with different numbers of vanishing elements
depending on the random initialization of all the variables involved.
In what follows,
we will explain in detail the application of these loss functions
and show numerical results for the obtained mass matrices and
the corresponding optimized parameters.

There are several possibilities for the counting functions with the behavior shown in
Eq.~\eqref{eq:fvcount} and~\eqref{eq:fnvcount}.
One simple form is given by
\begin{align}
    \label{eq:felemvc}
    f_\text{vc-s}(M_{jk};\epsilon_\text{vc},\alpha) & =  \frac{\epsilon_\text{vc}^\alpha}{\epsilon_\text{vc}^\alpha + \left| M_{jk}\right|^\alpha}\,, \\
    \label{eq:felemnvc}
    f_\text{nvc-s}(M_{jk};\epsilon_\text{nvc},\alpha) & = \frac{\left| M_{jk}\right|^\alpha}{\epsilon_\text{nvc}^\alpha + \left| M_{jk}\right|^\alpha}\,,
\end{align}
where the $\epsilon$ and $\alpha$ parameters
allow to fine tune the behavior of the functions,
affecting the performance of the optimization.
They can be decided beforehand,
and we will not vary them during optimization to improve its stability.
However, one could devise more complicated training
schedules that update, for example, the threshold based
on the updated values of the mass matrix elements.

\subsection{Random Baseline}

As a first test,
we randomly select the angles and phases in $V^R$
in random uniform distributions.
We limit the angles $\theta_{jk}^R$ to the range $[0, \pi/2]$
while for the phases we use the full range from $[-\pi,\pi]$.
From this first test, it is not expected that we will be able to say
anything new with regard to neutrino mass textures.
However, we obtain an interesting tendency with regard
to the element $M^\nu_{D11}$.
After running 5000 tests,
randomly selecting all the variables in Eq.~\eqref{eq:mnuundiag},
we find that $|M^\nu_{D11}|$
has a tendency to peak sharply around $\sim 0.01$~eV\@.

\begin{figure}[tb]
    \centering
    \includegraphics[scale=0.95]{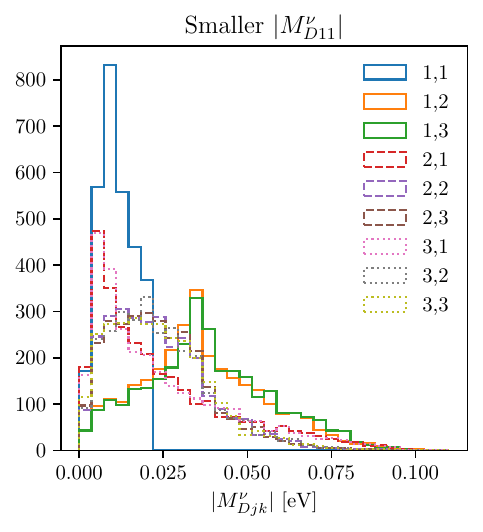}%
    \includegraphics[scale=0.95]{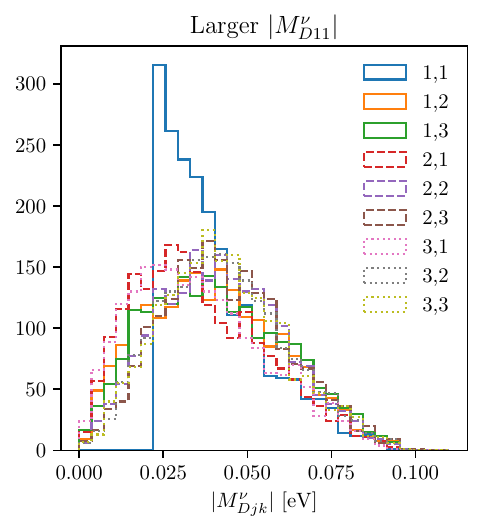}
    \caption{\label{fig:smalllarge11}%
        Distributions for the random test of sizes of $M^\nu_D$ entries.
        The distributions have been separated
        according to the size of $|M^\nu_{D11}|$,
        smaller $|M^\nu_{D11}|$ (left) is less than 0.022~eV
        while larger $|M^\nu_{D11}|$ (right) is above the same number.
    }
\end{figure}

In Figure~\ref{fig:smalllarge11}
we display distributions of elements
according to a hard cut applied on $|M^\nu_{D11}|$ at 0.022~eV\@.
In the left panel we show results around the peak,
i.e., $|M^\nu_{D11}| < 0.022$~eV, where
we find that elements in the same row,
$|M^\nu_{D12}|$ and $|M^\nu_{D13}|$,
have a tendency to exhibit sharp peaks
at around 0.035-0.040~eV\@.
Elements in the same column as $|M^\nu_{D11}|$, i.e., 
$|M^\nu_{D21}|$ and $|M^\nu_{D31}|$,
exhibit peaks at around half the position of the peak in $|M^\nu_{D11}|$.
Other elements do not feature sharp peaks
and follow similar distributions
with maximums close to 0.02~eV\@.
This already suggests the presence of relationships
among elements of $M^\nu_D$,
in particular for those in the same row
and column of $|M^\nu_{D11}|$
which present peaks at similar positions.
In the right panel of the same figure
we show distributions for samples to the right of the peak, i.e.,
$|M^\nu_{D11}| > 0.022$~eV\@.
We find that all the distributions for the other elements
tend to have similar shapes
with their maximums at approximately the same positions,
in the range 0.02-0.05 eV\@.
These distributions are roughly consistent with the sizes
of $|M^\nu_{D11}|$ shown in the same panel.
This could be taken as a hint of an approximately democratic texture
in the neutrino mass matrix.
In fact, the exactly democratic neutrino mass matrix can be diagonalized
by the exact form of the tribimaximal mixing,
ruled out by the non-zero $\theta_{13}$ observation.
Yet, it is interesting to see from this random test some coincidences
between the distributions of some elements,
considering that we have not assumed a particular
model other than Dirac neutrino masses
and we have used only random values.

\subsection{Learning Equality Between Two Entries}

\begin{figure}[tb]
    \centering
    \includegraphics[scale=0.95]{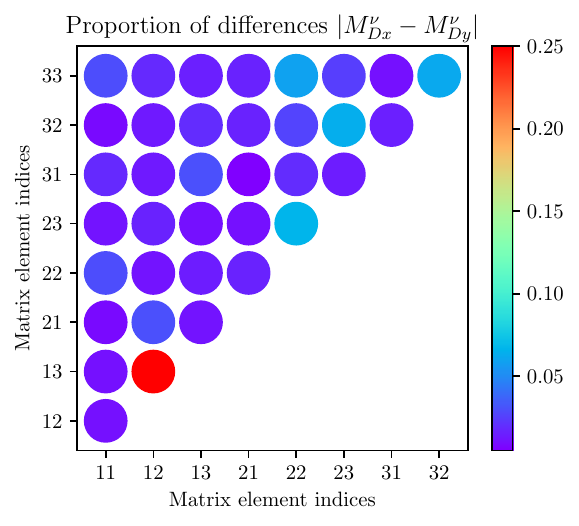}
    \caption{\label{fig:diffs_count}%
        Proportion of times that each pair of elements was choose
        for minimization in the loss function of Eq.~\eqref{eq:loss1diff}.
        Each circle represents a difference
        between the element with index in $x$ axis
        minus the element with index in $y$ axis.
        Only half is filled due to the symmetry
        $|M^\nu_{Djk} - M^\nu_{Dlm}| = |M^\nu_{Dlm} - M^\nu_{Djk}|$.
        The proportion represented by the color
        is the number of times the difference was chosen
        over the total number of tests that were used.
    }
\end{figure}

In this section, we proceed to actually attempt to \emph{learn}
the parameters of $V^R$ by minimizing a loss function
that includes only the absolute value
of the difference between two elements.
The loss function for this is given in Eq.~\eqref{eq:loss1diff}.
We sample neutrino oscillation parameters randomly
using gaussian distributions as described above.
The parameters of $V^R$ will be initialized randomly
from uniform distributions as indicated before.
In Fig.~\ref{fig:diffs_count} we show the proportion of times
that each pair of matrix elements has the minimum difference in $10,000$ random tests.
We see that the difference $|M^\nu_{D12} - M^\nu_{D13}|$
happens to be the most common one,
occurring almost 25\% of the times the test was run.
It is followed by
$|M^\nu_{D22} - M^\nu_{D23}|$,
$|M^\nu_{D32} - M^\nu_{D33}|$,
$|M^\nu_{D22} - M^\nu_{D33}|$,
and $|M^\nu_{D23} - M^\nu_{D32}|$.
The last two come at no surprise as they are related to the popular
$\mu$-$\tau$ symmetry~\cite{Fukuyama:1997ky,Ma:2001mr,Balaji:2001ex,Lam:2001fb,Harrison:2002et,Xing:2015fdg},
motivated by the form of tribimaximal mixing.
As mentioned before,
the values for the measured oscillation parameters are sampled from gaussian distributions
that fall inside the experimental range and are not included in the loss function.
From the resulting $M^\nu_D$ after initialization,
we select the minimum absolute difference between elements
and use it in the loss function of Eq.~\eqref{eq:loss1diff}.
To minimize the loss function,
we perform 1,000 optimization steps using the Adam~\cite{kingma2017adammethodstochasticoptimization} algorithm,
updating the parameters of $V^R$ at each step.
We use an initial learning rate of 0.01 and reduce it by a factor of 0.9 every 10 steps.
If the loss function remains above $10^{-6}$, we apply another 1,000 optimization steps
with a fixed learning rate of 0.001.
The final result is an $M^\nu_D$ matrix, where at least two elements
are equal down to several decimal places,
as well as the corresponding parameters for $V^R$.
We repeat this process (random initialization and minimization) for
each of the 5 most common cases,
1,000 times each,
and obtain values for the parameters contained in $V^R$
corresponding to before and after minimization.

\begin{figure}[tb]
    \centering
    \includegraphics[scale=0.8]{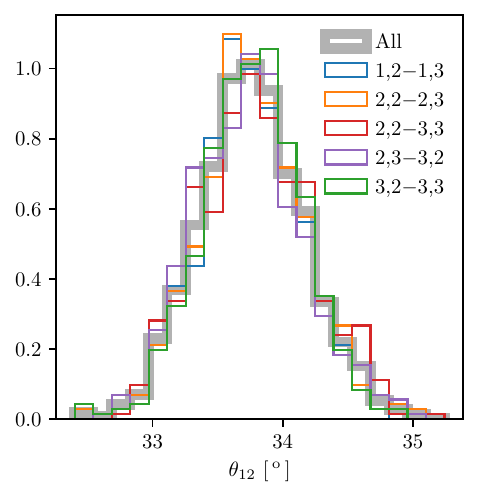}%
    \includegraphics[scale=0.8]{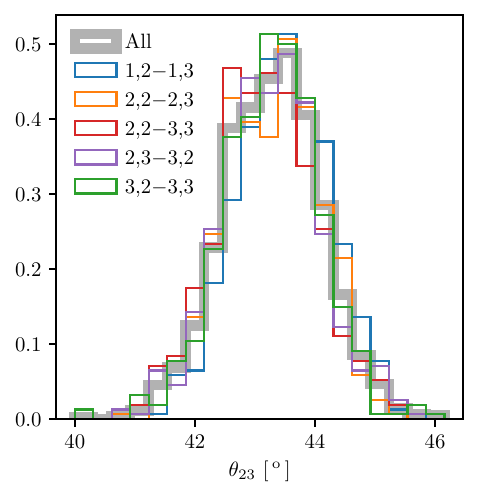}
    \includegraphics[scale=0.8]{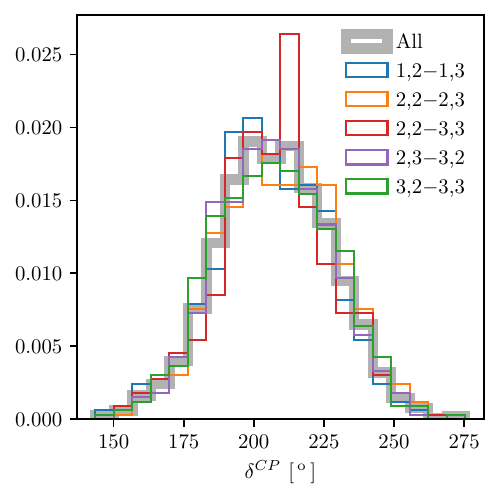}%
    \includegraphics[scale=0.8]{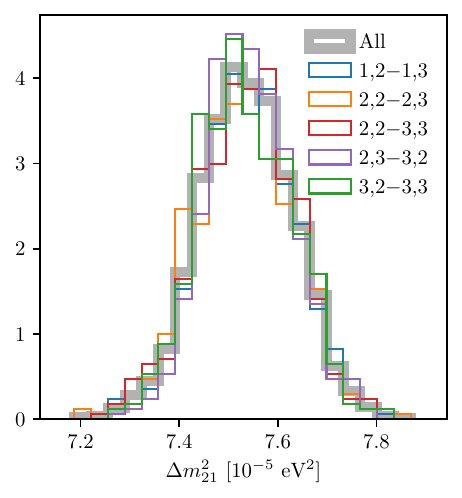}
    \caption{\label{fig:diffs_oscillation}%
        Distributions for the neutrino oscillation parameters for specific
        pairs of elements that were chosen for difference minimization.
        The gray histogram shows the distribution for all the cases.
        The colored histograms show the distribution only for selected
        differences.
    }
\end{figure}

Additionally, we check whether selecting specific differences has any effect
on the distributions of the neutrino oscillation parameters.
In Fig.~\ref{fig:diffs_oscillation},
we show the distributions of $\theta_{12}$, $\theta_{23}$,
$\delta^{CP}$, and $\Delta m_{21}^2$
for all the selected differences in color,
while the gray histogram shows the sampled distribution
for the corresponding parameters, regardless of the differences.
We see that colored and gray distributions follow similar
distributions,
indicating that selecting those particular differences
has almost no effect on the distribution of oscillation parameters.
The parameters that are not displayed, $\theta_{13}$ and $\Delta m_{31}^2$,
have been fixed to their central values.
Note that colored distributions can go above the gray distribution
because they have been normalized to compare their shapes.
Therefore, comparisons of height between the different distributions
are meaningless.

\begin{figure}[!htb]
    \centering
    \includegraphics[scale=0.78]{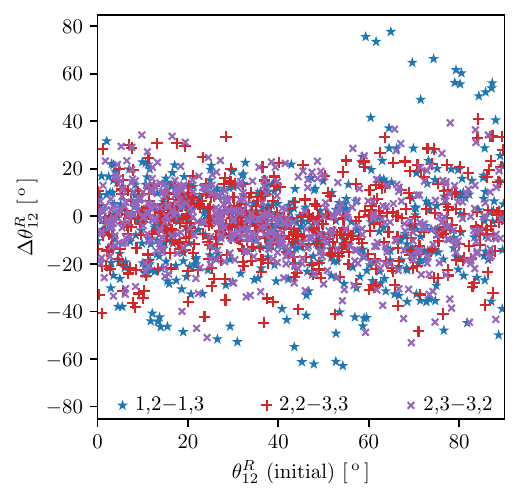}%
    \includegraphics[scale=0.78]{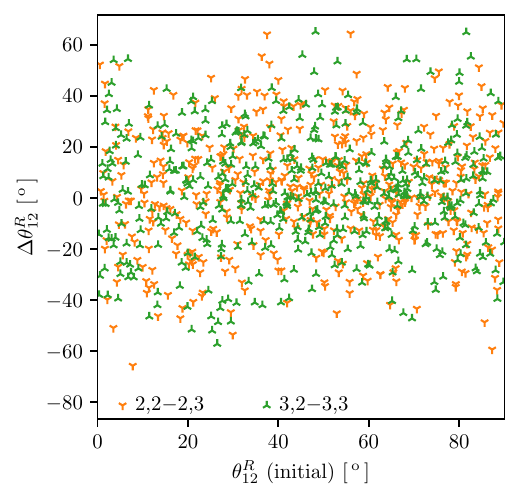}
    \includegraphics[scale=0.78]{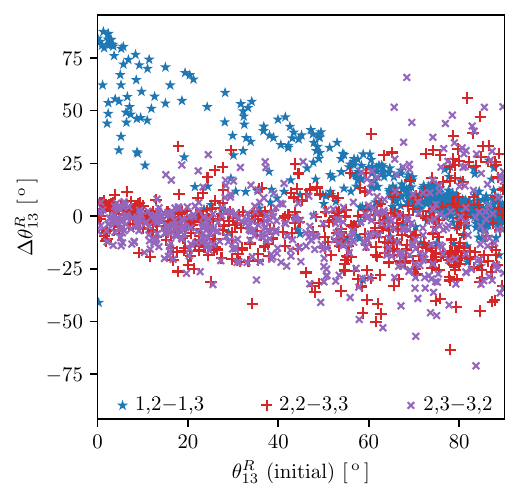}%
    \includegraphics[scale=0.78]{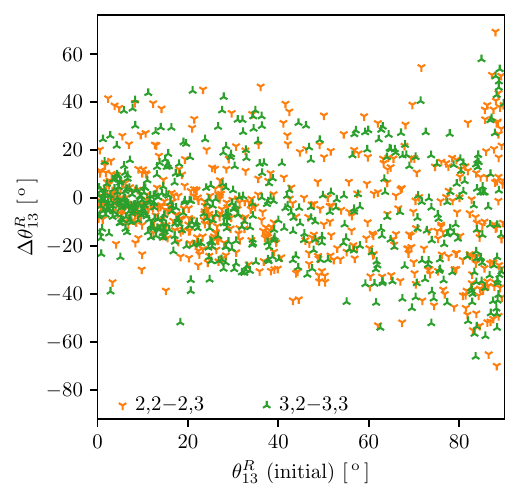}
    \includegraphics[scale=0.78]{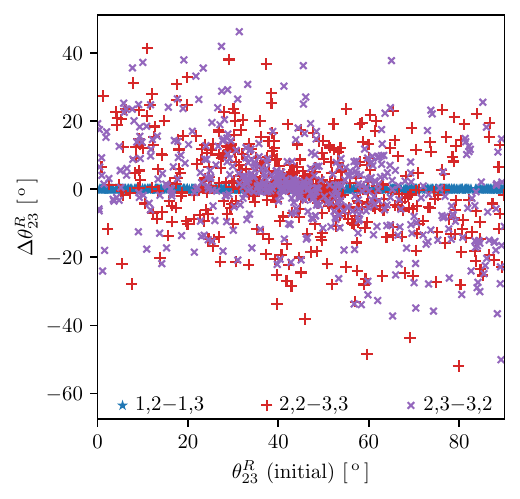}%
    \includegraphics[scale=0.78]{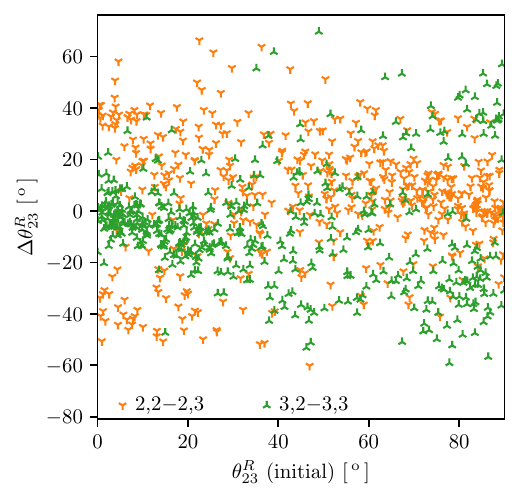}
    \caption{\label{fig:drifttheta}%
        Change in $V^R$ parameters after the minimization.
        The horizontal axis shows initial value, while the vertical axis
        shows the change in parameter value.
    }
\end{figure}

In the case of the parameters of $V^R$,
we expect to see a heavy dependence on the differences,
since these are the parameters that we have optimized
in the minimization process.
Namely, the parameters that have been included in the minimization process are
$\theta^R_{jk}$ ($jk=12,13,23$), $\delta^R$, $\phi^R_2$ and $\phi^R_3$.
In Fig.~\ref{fig:drifttheta},
we show the displacement of the $\theta^R_{jk}$ parameters
from their initial values,
with the initial values given in the $x$-axis.
On the $y$-axis, for each parameter, we use $\Delta$ to represent the difference between
the parameter value after optimization minus its initialization value, given on the $x$-axis.
In the case of the parameter $\theta^R_{12}$,
we see in the upper-left panel of Fig.~\ref{fig:drifttheta}
that the differences $|M^\nu_{D12} - M^\nu_{D13}|$,
$|M^\nu_{D22} - M^\nu_{D33}|$, and
$|M^\nu_{D23} - M^\nu_{D32}|$
tend to accumulate between $20\degree$ and $40\degree$,
with the displacement from minimization following
a rather random pattern.
The displacement for $|M^\nu_{D12} - M^\nu_{D13}|$
is much larger than for the other two differences in the same figure
and grows above the initial $60\degree$.
In the upper-right panel,
we see that $|M^\nu_{D22} - M^\nu_{D23}|$, and
$|M^\nu_{D32} - M^\nu_{D33}|$ have no preference for $\theta^R_{12}$
and that displacement from minimization is similar in both cases.
In the middle-left panel,
we see that $\theta^R_{13}$ accumulates around $90\degree$
for $|M^\nu_{D12} - M^\nu_{D13}|$,
while for $|M^\nu_{D22} - M^\nu_{D33}|$ and
$|M^\nu_{D23} - M^\nu_{D32}|$,
it tends to prefer to be closer to $0\degree$.
Interestingly,
for $|M^\nu_{D12} - M^\nu_{D13}|$,
minimization tends to take $\theta^R_{13}$
close to $90\degree$ while keeping it not exactly at that value,
resulting in the diagonal tendency that can be seen from initial $0\degree$ to $\sim 90\degree$
in the blue stars.
In the middle-right panel,
we see that $\theta^R_{13}$
is less sensitive to minimization
when it is initialized closer to $0\degree$
for both differences $|M^\nu_{D22} - M^\nu_{D23}|$ and
$|M^\nu_{D32} - M^\nu_{D33}|$.
In the lower panels, we display the displacement of $\theta^R_{23}$
as a function of its initial value.
In the lower-left panel we see that for the cases 
$|M^\nu_{D22} - M^\nu_{D33}|$ and
$|M^\nu_{D23} - M^\nu_{D32}|$
most initializations start around $45\degree$,
while the case
$|M^\nu_{D12} - M^\nu_{D13}|$
seems to be insensitive to the initial value of $\theta^R_{23}$
even showing no displacement after minimization.
We also see that optimization tends to displace $\theta^R_{23}$
upwards when initialized below $45\degree$
and downwards when initialized above it.
In the lower-right panel,
we see that the cases $|M^\nu_{D22} - M^\nu_{D23}|$
and
$|M^\nu_{D32} - M^\nu_{D33}|$
prefer to start at around $0\degree$ and $90\degree$,
respectively, with minimization having a rather random distribution.

\begin{figure}[!htb]
    \centering
    \includegraphics[scale=0.8]{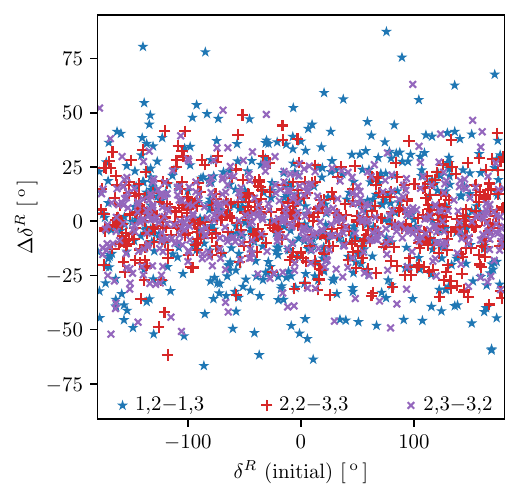}%
    \includegraphics[scale=0.8]{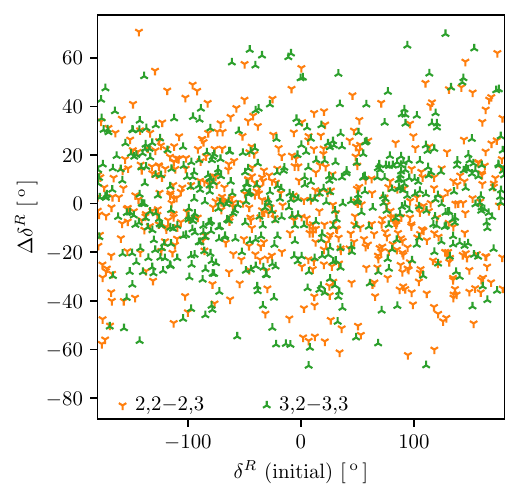}
    \includegraphics[scale=0.8]{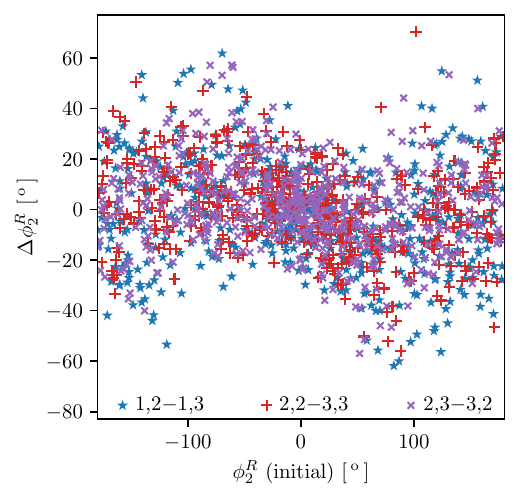}%
    \includegraphics[scale=0.8]{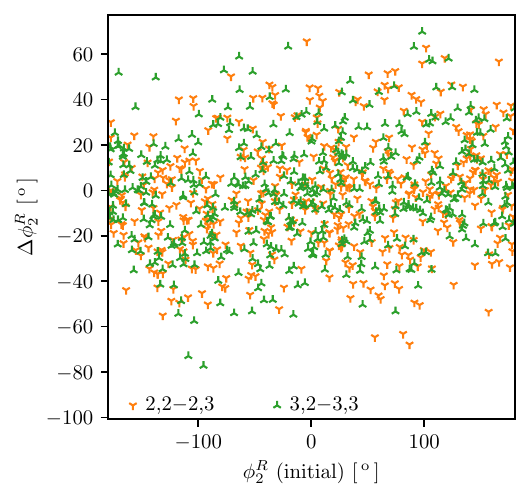}
    \includegraphics[scale=0.8]{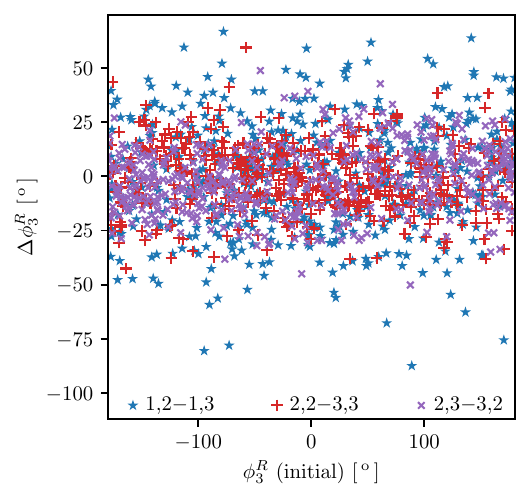}%
    \includegraphics[scale=0.8]{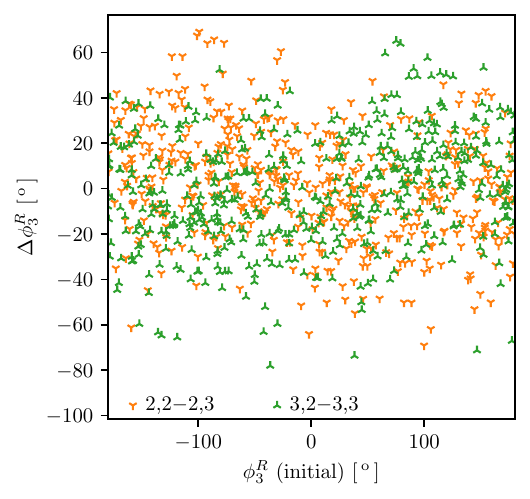}
    \caption{\label{fig:driftphases}%
        Change in $V^R$ parameters after the minimization.
        The horizontal axis shows initial value, while the vertical axis
        shows the change in parameter value.
    }
\end{figure}

Initializations and displacements during minimization
are shown for phases in Fig.~\ref{fig:driftphases}.
For this case, most initializations happen to be more evenly distributed,
and displacements also show little to no dependence on initialization phase.
However, a few interesting points can be made about initialization in this case.
For example, in the distribution of the initial phase for $\delta^R$,
as shown in the upper-right panel,
for $|M^\nu_{D32} - M^\nu_{D33}|$
there is a slight preference for initial values in the middle of the range between $0\degree$ and
$180\degree$ or $-180\degree$,
while for
$|M^\nu_{D22} - M^\nu_{D23}|$
this phase appear more uniformly distributed.
However, initialization of $\delta^R$ remains more evenly distributed when compared to distributions shown in Fig.~\ref{fig:drifttheta}.
Something similar can be said about $\phi^R_{3}$, for all the differences (bottom row),
and $\phi^R_{2}$ for $|M^\nu_{D22} - M^\nu_{D23}|$ and $|M^\nu_{D32} - M^\nu_{D33}|$ (middle right),
which seem to prefer initial values evenly distributed in the whole range
and displacements with no clear tendency.
One case with noticeable features is the distribution of displacements $\Delta \phi^R_2$
shown on the middle left.
Here we can see that for the three differences
$|M^\nu_{D12} - M^\nu_{D13}|$,
$|M^\nu_{D22} - M^\nu_{D33}|$,
and $|M^\nu_{D23} - M^\nu_{D32}|$,
optimization gives mostly positive displacements for $\Delta \phi^R_2$ when initialized below $0\degree$
and mostly negative displacements when it is initialized above $0\degree$.
This indicates that for these three differences optimization brings $\phi^R_2$ closer to $0\degree$.

In the following, we give a few numerical examples of $M_D^{\nu}$ after minimization.
These matrices contain two elements with the same complex value and can be
checked that they can be diagonalized as in Eq.~\eqref{eq:mnudiag},
with $U^\nu$ consistent with a $U_\text{PMNS}$ matrix
with oscillation parameters within experimental ranges.
After each matrix, we show the numerical results obtained for the physical mixing parameters and $m_{\nu}$ obtained from the values set during initialization (see the second paragraph of Sec.~\ref{sec:nulearn} and Fig.~\ref{fig:genchart}):
\begin{align}
    & \frac{M^\nu_D}{10^{-2}\text{~eV}} = \left(
    \begin{array}{ccc}
       0.42462+0.59258i & -3.35011+0.19065i & -3.35011+0.19065i \\
       -0.51059+0.07189i & -0.5556-0.08104i & 0.4283-0.22683i \\
       -0.06563+0.09068i & -0.0642-0.92411i & -0.29787-1.08555i
    \end{array}
    \right)\,, \\
    & s^2_{12} = 0.3114\,, \quad  s^2_{23} = 0.492\,, \quad  s^2_{13} = 0.02249\,, \quad
        \delta^{CP} = 238.47\degree\,, \quad J = -0.028926 \,, \nonumber\\
    & \Delta m^2_{21} = 7.574 \times 10^{-5} \text{ eV}^2\,, \quad \Delta m^2_{31} = 2.521 \times 10^{-3} \text{ eV}^2\,, \quad
        m_\nu = 8.9992\times 10^{-3} \text{ eV}\,, \nonumber \\
    & \frac{M^\nu_D}{10^{-2}\text{~eV}} = \left(
    \begin{array}{ccc}
       0.4842+0.91175i & -2.10139-5.06338i & -2.05571-5.10392i \\
       -5.96953+0.5551i & -0.51133+0.01792i & -0.51133+0.01792i \\
       0.06789+0.0167i & 4.18078+0.68614i & -4.26808-0.7743i
    \end{array}
    \right)\,, \\
    & s^2_{12} = 0.3041\,, \quad  s^2_{23} = 0.4753\,, \quad  s^2_{13} = 0.02249\,, \quad
        \delta^{CP} = 171.39\degree\,, \quad J = 0.0050392 \,, \nonumber\\
    & \Delta m^2_{21} = 7.541 \times 10^{-5} \text{ eV}^2\,, \quad \Delta m^2_{31} = 2.521 \times 10^{-3} \text{ eV}^2\,, \quad
        m_\nu = 6.0839\times 10^{-2} \text{ eV}\,, \nonumber \\
    & \frac{M^\nu_D}{10^{-2}\text{~eV}} = \left(
    \begin{array}{ccc}
       9.35176+0.04943i & 0.42776-0.5038i & 0.3971-0.52925i \\
       -0.0795-0.78652i & 8.79824+3.1097i & -0.66421+3.40956i \\
       -0.07241-0.83716i & -0.66421+3.40956i & 8.68342+3.69522i
    \end{array}
    \right)\,, \\
    & s^2_{12} = 0.3155\,, \quad  s^2_{23} = 0.4571\,, \quad  s^2_{13} = 0.02249\,, \quad
        \delta^{CP} = 207.02\degree\,, \quad J = -0.015416 \,, \nonumber\\
    & \Delta m^2_{21} = 7.507 \times 10^{-5} \text{ eV}^2\,, \quad \Delta m^2_{31} = 2.521 \times 10^{-3} \text{ eV}^2\,, \quad
        m_\nu = 9.4228\times 10^{-2} \text{ eV}\,. \nonumber
\end{align}

\subsection{Discovering $n$-zero Textures}
\label{sec:discoverzero}

Taking the previous section as a demonstration
of the general process of selection of texture and minimization,
now we proceed to apply the same process to a more complicated case, i.e., 
learning parameters for $n$-zero textures.
The study of such textures is popular in the literature~\cite{
    Fritzsch:2011qv,Meloni:2014yea,Nguyen:2014mwa,Benavides:2022hca,Ding:2022aoe,Devi:2023vpe,Rico:2023drr,
    Zhao:2023vkb,Kumar:2023iaj,Zhou:2024gcp,Nomura:2023usj,Carrillo-Monteverde:2024ntq,Singh:2024ivf,
    Mazumder:2024lne,Kobayashi:2024cvp,Kobayashi:2025znw,Kobayashi:2025ldi,Calibbi:2025ded,Biswas:2025woq,Gogoi:2025htb,Jiang:2025psz,Borah:2025vtn,Priya:2026tpk
    }
as they can be obtained from minimal models with a reduced number of parameters.
Thus, they represent predictive restricted scenarios.
Such textures typically are obtained from hypothesizing
a combination of new symmetries and fields
that result in vanishing entries in the neutrino mass matrix.
Note that we do not expect the optimization process to
determine elements with exactly vanishing values,
but instead we will consider the matrices which are found to have
$n$ elements much smaller than the rest to be the $n$-zero candidates.
To determine if a particular matrix is a successful candidate,
we will set its $n$ smaller elements to exactly vanish
and recalculate the corresponding predictions
for the numbers in Table~\ref{tab:measurements}
using Eqs.~\eqref{eq:VRfromsquaremnu}--\eqref{eq:mnukatrin}.
In the two methods presented next,
we will initialize oscillation parameters using random gaussian distributions,
as described in Sec.~\ref{sec:nulearn},
but we will keep only initializations within 2$\sigma$ of those gaussian distributions.
Then, for the final matrices with $n$ elements set to zero,
we will check if the predictions remain within the 2$\sigma$ range
cited in Table~\ref{tab:measurements}.
The matrices that stay within that range will be marked as surviving $n$-zero textures.
The remainder of the details for each method will be described in the next section.

\subsubsection{Discovering $n$-zero Textures by Random Search Plus Optimization}
\label{sec:nzerorandom}

In this section, we describe our first approach to find zero textures that are hinted at by
the measured oscillation parameters.
The optimization process will roughly follow the same process as in the previous section but
with a loss function specific to this section.
We will make a few assumptions during our search:
\begin{itemize}
    \item
        The number of vanishing entries will be no more than five.
    \item
        No column or row will be full of vanishing entries.
\end{itemize}
The steps we will follow to consider that oscillation parameters hint at a particular $n$-zero texture
are as follows:
\begin{enumerate}
    \item
         Sample randomly from oscillation parameters using gaussian distributions
         as in the previous section, but only within 2$\sigma$ ranges.
         The parameters of $V^R$ and lightest neutrino mass will be sampled from a random uniform
         distribution.
    \item
        Use Eqs.~\eqref{eq:mnuundiag},~\eqref{eq:mnu2} and~\eqref{eq:mnu3} to obtain numerical values for $M^\nu_D$.
    \item
        Sort the elements of $|M^\nu_D|$ from smallest to largest and take
        ratios sequentially, $|M^\nu_{Dab}|/|M^\nu_{Dcd}|$, where $|M^\nu_{Dcd}|$
        is the value smaller than $|M^\nu_{Dab}|$ but closest to it.
        This will result in eight ratios characterizing the change in size between elements
        that are closer in magnitude.
    \item
        Select the largest ratio, $\max(|M^\nu_{Dab}|/|M^\nu_{Dcd}|)$, representing the biggest change
        in size between elements, as the division between elements that will vanish and elements that do not.
        For example, after determining $\max(|M^\nu_{Dab}|/|M^\nu_{Dcd}|)$ for a particular $ab$ entry,
        all the entries with magnitude below $|M^\nu_{Dab}|$ will be considered as the elements that vanish.
    \item
        Construct a loss function that minimizes the magnitude of the entries that were selected in the previous step.
    \item
        Apply minimization as was done in the previous section.
\end{enumerate}
The loss function employed in this approach is given in Eq.~\eqref{eq:lossnzero} and
constructed with the set of elements chosen in Step 4 of the process described above
as vanishing elements.
This loss function works by separating the values of the average of the two sets of elements,
making the numerator small and the denominator large relative to each other.
For this case, we will also use the lightest neutrino mass as part of the minimization.
Therefore, we will also include a term for KATRIN's measurement of the effective electron-antineutrino mass as
given in Eq.~\eqref{eq:losskatrin}, with the weight $w_K = 1000$~eV$^{-2}$.

\begin{figure}[tpb]
    \centering
    \includegraphics[width=0.4\textwidth]{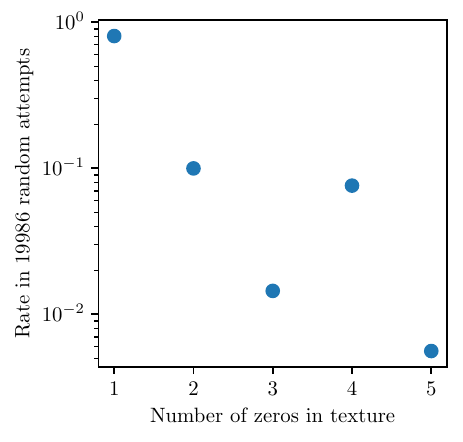}%
    \caption{\label{fig:randomrate}%
        Rate of found textures by number of zeros by using the random search described in Sec.~\ref{sec:discoverzero}.
    }
\end{figure}

Given the process described above,
the rate of finding textures with each number of zeros
is shown in Fig.~\ref{fig:randomrate}.
As expected, in this figure we can see that it is easier to find
1-zero textures.  Additionally, we see that the rate goes down as we search for more zeros,
except in the case of 4-zero textures where the rate is comparable to 2-zeros.
To try to quantify the minimizing power of the process above,
we search randomly for $n$-zero textures as described above
and collect a total of 500 examples per $n$-zero texture,
with $n=1,2,3,4,5$.
Initially, we minimize the loss function during 1,000 optimization steps,
attempting 1,000 more steps in cases where the loss function of Eq.~\eqref{eq:lossnzero}
does not go below $10^{-18}$.
For the optimization, we use the Adam algorithm
with an initial learning rate of 0.001
reduced by a factor of 0.99 every 10 steps.

\begin{figure}[!htb]
    \centering
    \includegraphics[scale=0.7]{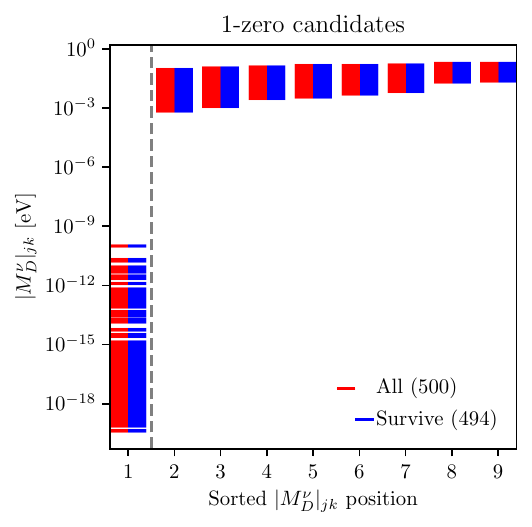}%
    \includegraphics[scale=0.7]{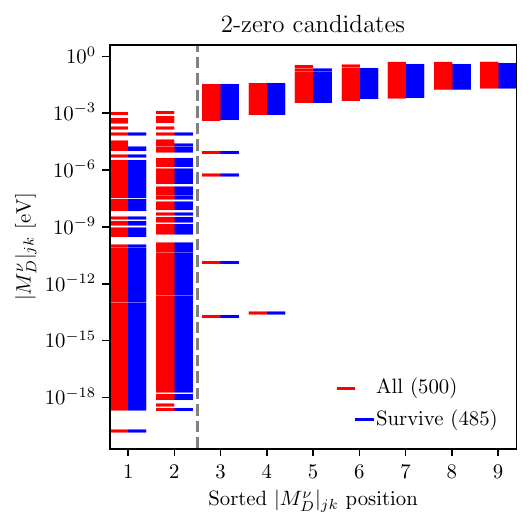}
    \includegraphics[scale=0.7]{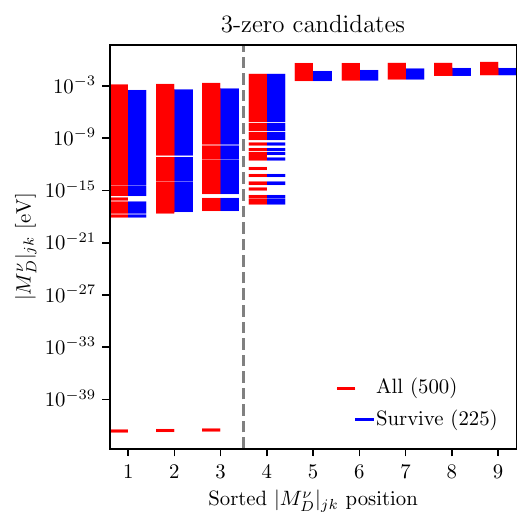}%
    \includegraphics[scale=0.7]{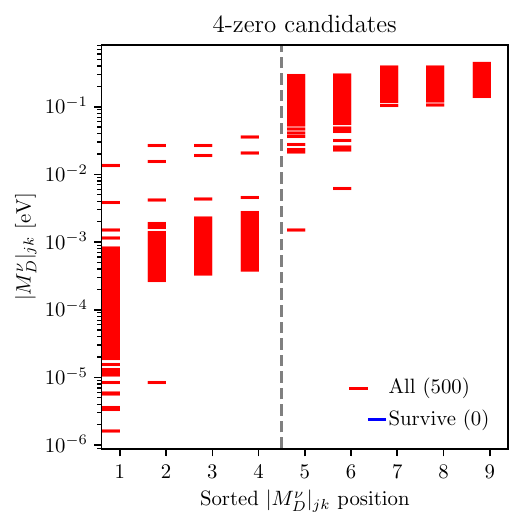}
    \includegraphics[scale=0.7]{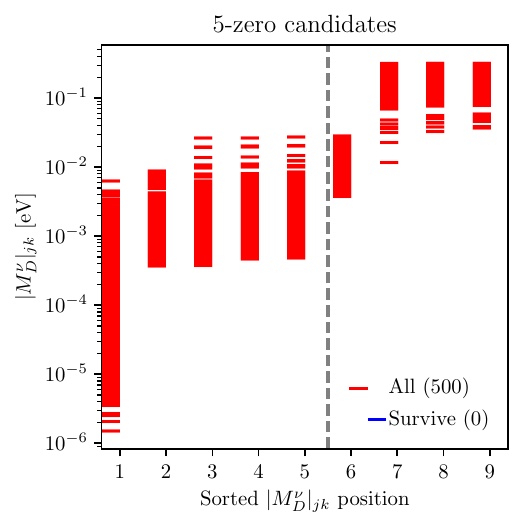}%
    \caption{\label{fig:patternsrandom}%
        Sorted absolute values of elements for the optimized matrices
        using random search as described in Sec.~\ref{sec:nzerorandom}.
        The red bars show all the obtained $n$-zero texture candidates.
        The blue bars correspond to matrices where the $n$ smallest absolute values
        have been set to zero while still obtaining all oscillation parameters
        inside their 2$\sigma$ ranges (labelled ``Survive'').
        The number in parenthesis shows counts of $M^{\nu}_D$
        obtained in the process.
    }
\end{figure}

\begin{figure}[htb]
    \centering
    \includegraphics[width=0.20\textwidth]{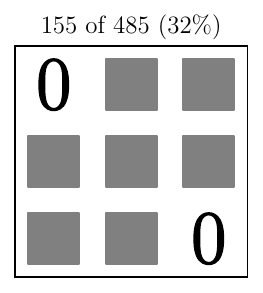}%
    \includegraphics[width=0.20\textwidth]{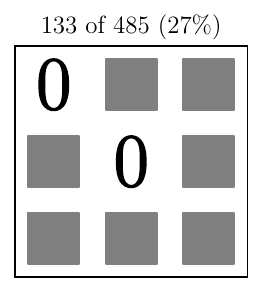}%
    \includegraphics[width=0.20\textwidth]{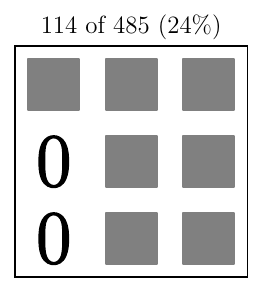}
    \includegraphics[width=0.20\textwidth]{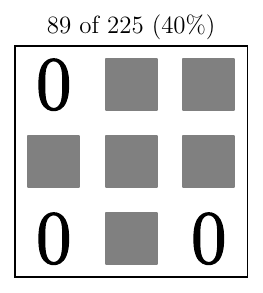}%
    \includegraphics[width=0.20\textwidth]{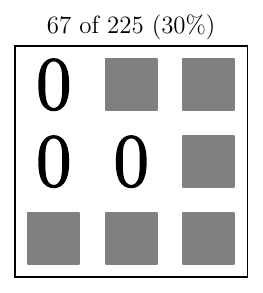}%
    \includegraphics[width=0.20\textwidth]{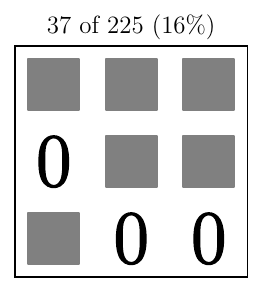}
    \caption{
        The three most commonly found $n$-zero textures
        using the random search algorithm of Sec.~\ref{sec:nzerorandom}
        for $n=2$ (top) and 3 (bottom).
        The counts above each figure represent the number of times the texture occurs
        in the surviving cases of Fig.~\ref{fig:patternsrandom}.
        These counts also consider equivalent textures due to row permutations.
        The figures are sorted from most common on the left and descending to the right.
        The gray squares correspond to values shown in Fig.~\ref{fig:patternsrandom} 
        to the right of the gray dashed vertical line
        (2-zero and 3-zero cases).
    }
    \label{fig:commonrandom}
\end{figure}

As previously discussed,
we do not expect the optimization process to deliver
matrices with exactly $n$ elements equal to zero but instead to deliver $n$ elements much smaller than the others.
These optimized matrices will be considered $n$-zero texture candidates.
Then, we do a final test by setting the $n$ smallest elements to zero
and checking if the resulting matrix still follows experimental constraints.
In particular, we require that the oscillation parameters
remain within 2$\sigma$ of the measurements cited in Table~\ref{tab:measurements}
and that the value of $m_\nu$ remains below 0.45~eV,
the upper limit set by KATRIN~\cite{KATRIN:2024cdt}.
The patterns of absolute values for the elements of $M^\nu_D$
obtained by the optimization process described here
are shown in Fig.~\ref{fig:patternsrandom}.
The elements have been sorted according to their absolute values
which are shown in increasing order from left to right.
The red bars show results for all optimized matrices,
while the blue bars show results only for the matrices
that survived the experimental tests
after setting the smallest $n$ elements to zero.
The dashed gray line separates the elements that were set to zero
from those that were kept at their optimized values.
The numbers in parentheses in the legend
indicate the number of displayed results corresponding to that label.
For 1-zero candidates (upper left) and 2-zero candidates (upper right),
we see that minimization worked as expected delivering one and two
much smaller elements, respectively.
For these two cases, we also get a survival of $\sim490$ out of 500 optimized
matrices.
In the case of 3-zero candidates (middle left),
we see that the minimization performance is highly effective,
sometimes delivering more than three substantially smaller elements.
This is related to cases where the optimization attempts to minimize two elements in
the same row resulting in accidental minimization of the remaining element.
Survival in this case is lower with 225 cases passing the applied tests.
For 4-zero candidates (middle right) and 5-zero candidates (lower),
the optimization was much less efficient,
delivering inconsistent minimizations and, moreover,
resulting in no surviving cases.
One particular characteristic of the 4- and 5-zero optimizations
is that a single element seems to have been minimized more efficiently than the rest.
The three most commonly surviving textures found using the algorithm described here
are shown in Fig.~\ref{fig:commonrandom}
for $n=2$ and 3.
Counts for the number of times each texture occurs are shown above
each case, counting also equivalent textures that are row permutations
of the displayed cases.

Some examples of numerical matrices
for 2- and 3-zeros are given below,
including their corresponding values for observables.
The numerical values for the mixing observables and $m_{\nu}$ shown below each matrix have been obtained
using Eqs.~\eqref{eq:VRfromsquaremnu}--\eqref{eq:mnukatrin}.
In the case of $\delta^{CP}$,
there is an ambiguity in the quadrant due to the range of the arcsine function.
To solve this ambiguity,
we assume that the predicted $\delta^{CP}$ stays close to the initialization range,
which in this section corresponds to [167\degree,247\degree].
Therefore, we take $\delta^{CP} = \pi - \sin^{-1}[\ldots]$,
where $\sin^{-1}[\ldots]$ corresponds to the right-hand side of Eq.~\eqref{eq:arcsinejarlskog}, to arrive at the following numerical matrices and observables:
\begin{align}
        & \frac{M^\nu_D}{10^{-2}\text{~eV}} = \left(
    \begin{array}{ccc}
       0.74834+0.25015i & -3.46065+0.12821i & -3.6963+0.18086i \\
       -0.80448-0.80651i & -0.27798-0.31313i & 0 \\
       0 & -0.71088+0.50896i & 0.63125-0.6378i
    \end{array}
    \right)\,, \\
    & s^2_{12} = 0.3025\,, \quad  s^2_{23} = 0.4656\,, \quad  s^2_{13} = 0.02249\,, \quad
        \delta^{CP} = 203.04\degree\,, \quad J = -0.013145 \,, \nonumber\\
    & \Delta m^2_{21} = 7.373 \times 10^{-5} \text{ eV}^2\,, \quad \Delta m^2_{31} = 2.521 \times 10^{-3} \text{ eV}^2\,, \quad
        m_\nu = 1.3857\times 10^{-2} \text{ eV}\,, \nonumber \\
    & \frac{M^\nu_D}{10^{-2}\text{~eV}} = \left(
    \begin{array}{ccc}
       1.90766 & -1.02225+0.59142i & -1.40581+0.6545i \\
       0 & 2.35306+0.01696i & 0 \\
       0 & 2.70405+0.04954i & 3.76622
    \end{array}
    \right)\,, \\
    & s^2_{12} = 0.3195\,, \quad  s^2_{23} = 0.4495\,, \quad  s^2_{13} = 0.02249\,, \quad
        \delta^{CP} = 207.04\degree\,, \quad J = -0.015457 \,, \nonumber\\
    & \Delta m^2_{21} = 7.616 \times 10^{-5} \text{ eV}^2\,, \quad \Delta m^2_{31} = 2.521 \times 10^{-3} \text{ eV}^2\,, \quad
        m_\nu = 1.9077\times 10^{-2} \text{ eV}\,. \nonumber
\end{align}

\noindent Now that we have given examples of numerical matrices for 2- and 3-zero textures as well as their predictions, we proceed to explore how to discover $n$-zero textures via optimization.

\subsubsection{Discovering $n$-zero Textures Via Optimization}
\label{sec:nzerolosssearch}

The idea of this section is to use the optimization process of minimizing the
loss function to actually identify elements in the neutrino mass matrix that
could be considerably smaller than others.
We do not expect that minimization itself will make them completely vanish,
but we will be comfortable with them being several orders of magnitude smaller
such that setting them to zero makes little to no difference.
We will test the two loss functions given in Eqs.~\eqref{eq:lossvcount} and~\eqref{eq:lossnvcount},
to determine if there is any difference in behavior when
counting vanishing or non-vanishing elements, respectively.

The strategy to find textures and the parameters that characterize them is the following:
\begin{enumerate}
    \item
        Randomly initialize all the parameters as in the previous cases.
    \item
        Optimize the parameters using the loss function of either Eq.~\eqref{eq:lossvcount} or~\eqref{eq:lossnvcount}
        to determine which elements can be counted as vanishing.
        This can be done by selecting elements that were counted as vanishing in the last optimization step.
    \item
        Attempt to fully minimize the selected elements as much as possible if they have not reached
        the point of being vanishingly small when compared to other elements.
    \item
        Count the actual number of vanishing elements depending on the outcome of this second minimization.
\end{enumerate}
The minimization in point number 3 above is done with a simple loss function with the elements that were selected as vanishing
from the first minimization.

\begin{figure}[!htb]
    \centering
    \includegraphics[width=0.5\textwidth]{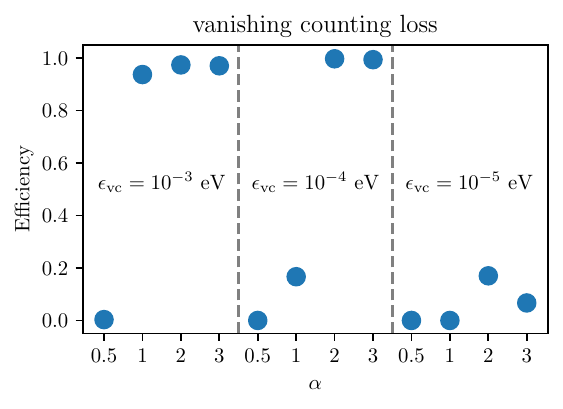}%
    \includegraphics[width=0.5\textwidth]{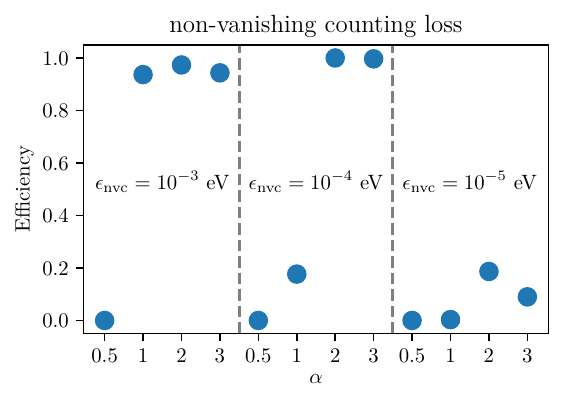}
    \includegraphics[width=0.5\textwidth]{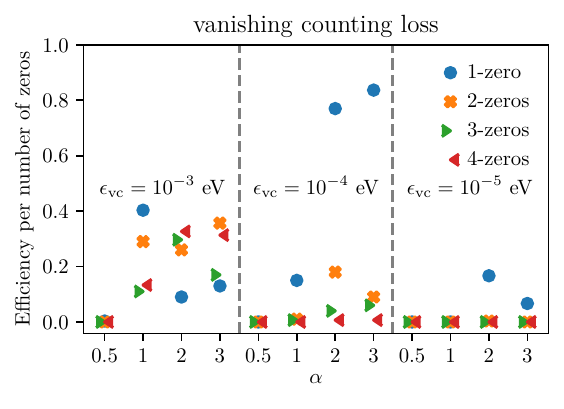}%
    \includegraphics[width=0.5\textwidth]{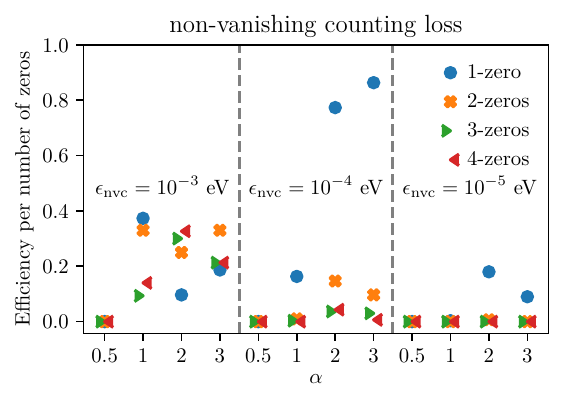}
    \caption{\label{fig:vcnvceff}%
        Efficiencies for the element counting loss functions.
    }
\end{figure}

We start by estimating good values for the loss function parameters $\epsilon_\text{vc,nvc}$ and $\alpha$.
To do this,
we perform 300 minimizations
of the loss functions in Eqs.~\eqref{eq:lossvcount} and~\eqref{eq:lossnvcount},
with individual matrix element losses defined by Eqs.~\eqref{eq:felemvc} and~\eqref{eq:felemnvc}, respectively.
We use all combinations of the values $\epsilon_\text{vc,nvc}/\text{eV} \in \{10^{-3}, 10^{-4}, 10^{-5}\}$
and $\alpha \in \{0.5, 1, 2, 3\}$.
Considering that minimizing $n$ elements of a complex matrix
is equivalent to attempting to bring $n$ real parts and $n$ imaginary parts down to zero,
using 7 unconstrained parameters (three $\theta_{jk}^R$, two $\phi_{2,3}^R$, $\delta^R$ and $m_1$)
may not achieve great results for $n > 3$.
Optimistically,
we fix $n_\text{max} = 5$ in Eq.~\eqref{eq:lossvcount},
which means that we will let the search go as far as 5-zero textures.
Our objective is to obtain efficiencies
that indicate how many times the optimization process
results in hints of textures,
always starting from randomly initializing all the variables in Eq.~\eqref{eq:mnuundiag}.
Efficiencies are simply calculated as (number of textures found)/(number of optimizations performed).
To consider that an element $M_{jk}$ has been ``chosen'' as vanishing by the loss function,
we set a threshold of $f_\text{vc} (M_{jk}) > 0.99$ ($f_\text{nvc} (M_{jk}) < 0.01$).
The resulting efficiencies from this test are shown in Fig.~\ref{fig:vcnvceff}.
Interestingly, when considering global efficiency (finding at least 1-zero texture),
shown in the two upper panels of Fig.~\ref{fig:vcnvceff},
we find that efficiencies have very similar values
across all the values tested for $\epsilon_\text{vc,nvc}$ and $\alpha$.
We also note that,
as $\epsilon_\text{vc,nvc}$ decreases,
it becomes difficult to reach an efficiency above 0.5.
The highest efficiencies are achieved with $\epsilon_\text{vc,nvc} = 10^{-4}$~eV,
for $\alpha = 2, 3$.
However, using $\epsilon_\text{vc,nvc} = 10^{-4}$~eV results in high global efficiency
for $\alpha \in \{1, 2, 3\}$.
Using $\epsilon_\text{vc,nvc} = 10^{-5}$~eV gives the worst performance for all the values
of $\alpha$ considered.
When looking at efficiencies for individual textures,
shown in the two bottom panels of Fig.~\ref{fig:vcnvceff},
we see that for $\epsilon_\text{vc,nvc} \in \{10^{-4}$~eV$,10^{-5}$~eV\},
1-zero textures dominate over the other textures,
while a more even distribution of textures is found for $\epsilon_\text{vc,nvc} = 10^{-3}$,
at least for $\alpha \in \{1, 2, 3\}$.
Notably, for $\alpha \in \{2, 3\}$
more optimizations resulted in textures with more than one zero.
It is precisely here where the vanishing and non-vanishing counting loss functions
show some difference,
as, apparently, the non-vanishing counting loss with $\epsilon_\text{vnc} = 10^{-3}$~eV
and $\alpha = 3$ seems to have a more even distribution of efficiencies,
with all 1-to-4-zero textures in the range 0.2--0.4.
Expectedly, none of the combinations of $\epsilon_\text{vc,nvc}$ and $\alpha$,
gave a loss function that achieved any textures with more than 4-zeros.
Therefore, 5-zero textures and above should have an efficiency below $1/300$.
Again, this is closely related to using only 7 free parameters during optimization.

\begin{figure}[!htb]
    \centering
    \includegraphics[scale=0.7]{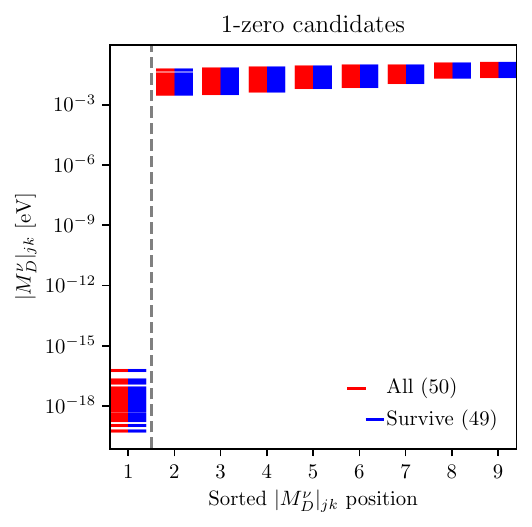}%
    \includegraphics[scale=0.7]{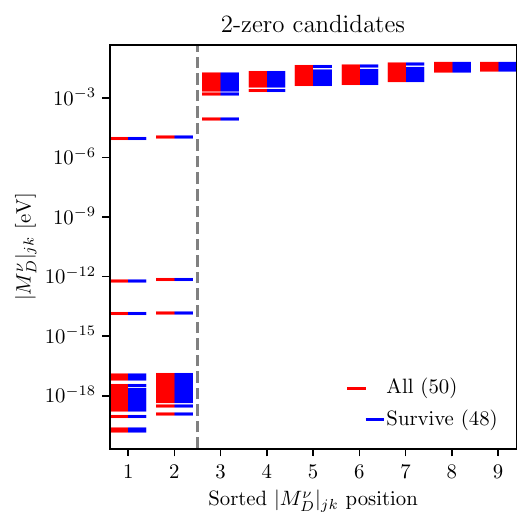}
    \includegraphics[scale=0.7]{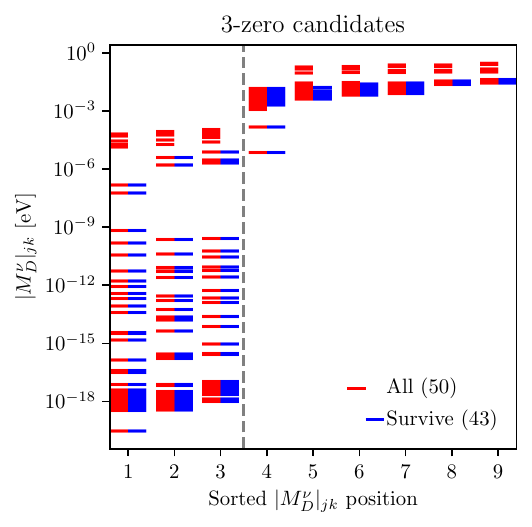}%
    \includegraphics[scale=0.7]{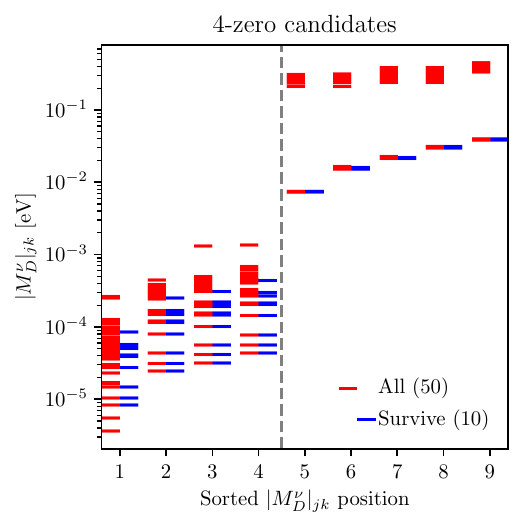}
    \caption{\label{fig:patternlosssearch}%
        Sorted absolute values of elements for the optimized matrices
        using non-vanishing counting loss function search
        as described in Sec.~\ref{sec:nzerolosssearch}.
        The red bars show all the obtained $n$-zero texture candidates.
        The blue bars correspond to matrices where the $n$ smallest absolute values
        have been set to zero while still obtaining all oscillation parameters
        inside their 2$\sigma$ ranges (labeled ``Survive'').
        The number in parenthesis shows counts of $M^{\nu}_D$
        obtained in the process.
    }
\end{figure}

\begin{figure}[htb]
    \centering
    \includegraphics[width=0.20\textwidth]{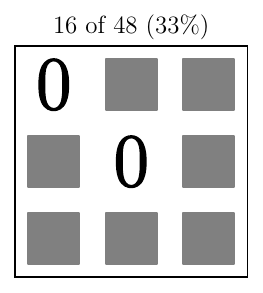}%
    \includegraphics[width=0.20\textwidth]{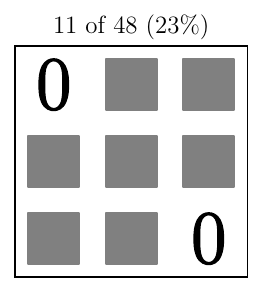}%
    \includegraphics[width=0.20\textwidth]{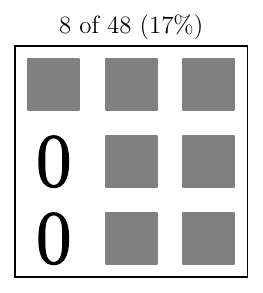}
    \includegraphics[width=0.20\textwidth]{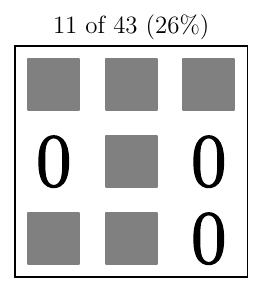}%
    \includegraphics[width=0.20\textwidth]{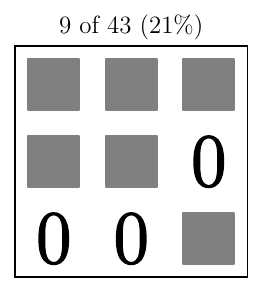}%
    \includegraphics[width=0.20\textwidth]{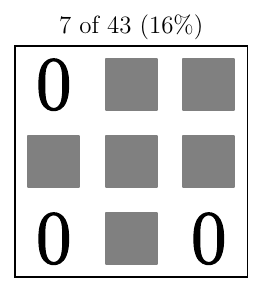}\\
    \includegraphics[width=0.20\textwidth]{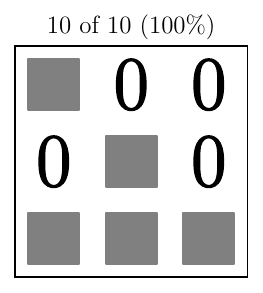}
    \caption{
        The three most commonly found $n$-zero textures
        using the optimizer search algorithm of Sec.~\ref{sec:nzerorandom}
        for $n=2$ (top) and 3 (middle).
        For $n=4$, all cases found are consistent with the texture shown in the bottom.
        The counts above each figure represent the number of times the texture occurs
        in the surviving cases of Fig.~\ref{fig:patternlosssearch}.
        These counts also consider equivalent textures due to row permutations.
        The figures are sorted from most common on the left and descending to the right.
        The gray squares correspond to values shown in Fig.~\ref{fig:patternlosssearch} 
        to the right of the gray dashed vertical line
        (2-zero, 3-zero and 4-zero cases).
    }
    \label{fig:commonlosssearch}
\end{figure}

Considering the information gained on efficiency per texture,
we perform a dedicated search using the non-vanishing counting loss
with parameters set at $\epsilon_\text{nvc} = 10^{-3}$~eV
and $\alpha = 3$.
We attempt a search of $n$-zero textures with $n\leq 4$,
by performing multiple optimizations and collecting results
after 1,000 steps.
The optimization algorithm is Adam with a fixed learning rate of 0.001.
At the end of the 1,000 optimization steps,
we determine which elements have absolute values
much smaller than the rest and use that information to count
the resulting number of vanishing elements.
In particular, we require that the value of $f_\text{nvc-s}((M^\nu_D)_{jk})$
of Eq.~\eqref{eq:felemnvc}
is below a threshold of 0.01 to consider that the $(M^\nu_D)_{jk}$ element could vanish.
After that,
we attempt to minimize the chosen elements again
but this time with a loss function of the type given in Eq.~\eqref{eq:lossnzero}.
This second optimization is analogous to the one of Sec.~\ref{sec:nzerorandom},
but with an initial learning rate of 0.001 and a reduction by a factor of 0.99 every 10 steps.
As before, we require that all the found textures
have at least one non-zero element per column and row.
This search takes longer, and therefore we only attempt to collect
50 candidates per texture.
The sorted absolute values for the collected candidates are shown in Fig.~\ref{fig:patternlosssearch},
including the results for the surviving cases.
As in the previous approach,
we set the $n$ elements of $M^\nu_D$ with the smallest absolute values to zero
and use Eqs.~\eqref{eq:VRfromsquaremnu}--\eqref{eq:mnukatrin}
to obtain the corresponding observables.
We consider the cases where the observables remain inside 2$\sigma$ ranges
as surviving cases, shown as blue bars in Fig.~\ref{fig:patternlosssearch}.
For 1- and 2-zero candidates, shown in the upper-left and upper-right panels, respectively,
the result is somewhat compatible with those shown
in Fig.~\ref{fig:patternsrandom}.
The sizes of sorted elements is similar in both figures,
and the survival rates have a similar size
when considering the ``Survive'' to ``All'' ratio.
However, considerable differences appear for 3- and 4-zero candidates,
shown in the lower-left and lower-right panels, respectively.
The survival rates (as ``Survive'' to ``All'' ratio) are larger
than in the corresponding cases in Fig.~\ref{fig:patternsrandom},
most notably for 4-zero candidates where a few cases were found.
In the case of 3-zero candidates,
the surviving cases, indeed, have 3 elements that are much smaller
than the other 6 as one would expect.
Similarly, for 4-zero candidates,
all the textures collected have 4 elements always smaller than the other 5.
The surviving cases have a narrower variation for the elements to the right of the gray dashed line,
when compared to the other 1-, 2- and 3-zero candidates,
but this could be related to the low surviving rate.
For all 4 panels shown in Fig.~\ref{fig:patternlosssearch},
we see that optimization using the non-vanishing-counting loss function
of Eq.~\eqref{eq:lossnvcount}
gives results that most of the time consistently minimize
only the indicated number of elements.
The reason for this is that this type of loss function
identifies the number of elements being minimized
and, at the end of optimization, reports the number of elements
that were not minimized.
The three most commonly surviving textures found using the algorithm described here
are shown in Fig.~\ref{fig:commonlosssearch}
for $n=2$ and 3.
In the case of $n=4$, all the surviving textures are permutations of the
textures displayed in the bottom row of Fig.~\ref{fig:commonlosssearch}.
Counts for the number of times each texture occurs are shown above
each case, counting also equivalent textures that are row permutations
of the cases displayed.

In the following, we show examples of surviving matrices with 2-, 3- and 4-zero textures
with their corresponding
values for the mixing angles and $m_{\nu}$, obtained
in the same way as in Sec.~\ref{sec:nzerorandom},
with the same solution for the ambiguity of $\delta^{CP}$:
\begin{align}
    & \frac{M^\nu_D}{10^{-2}\text{~eV}} = \left(
    \begin{array}{ccc}
       0.5375+0.47302i & -2.71315-1.77586i & -2.77915-1.8634i \\
       0.37786-0.38321i & 0 & -0.62809+0.71107i \\
       0 & -0.89509+0.94085i & -0.57059+0.74175i
    \end{array}
    \right)\,, \\
    & s^2_{12} = 0.3143\,, \quad  s^2_{23} = 0.4819\,, \quad  s^2_{13} = 0.02249\,, \quad
        \delta^{CP} = 187.35\degree\,, \quad J = -0.0043512 \,, \nonumber\\
    & \Delta m^2_{21} = 7.588 \times 10^{-5} \text{ eV}^2\,, \quad \Delta m^2_{31} = 2.521 \times 10^{-3} \text{ eV}^2\,, \quad
        m_\nu = 8.9570\times 10^{-3} \text{ eV}\,, \nonumber\\
    & \frac{M^\nu_D}{10^{-2}\text{~eV}} = \left(
    \begin{array}{ccc}
       -0.55823-0.53543i & 3.30143-0.56209i & 3.54834-0.61742i \\
       -0.5523+0.05066i & -0.75365-0.12562i & 0 \\
       0 & -0.74915-0.24974i & 0
    \end{array}
    \right)\,, \\
    & s^2_{12} = 0.3074\,, \quad  s^2_{23} = 0.4875\,, \quad  s^2_{13} = 0.02249\,, \quad
        \delta^{CP} = 237.83\degree\,, \quad J = -0.02862 \,, \nonumber\\
    & \Delta m^2_{21} = 7.49 \times 10^{-5} \text{ eV}^2\,, \quad \Delta m^2_{31} = 2.521 \times 10^{-3} \text{ eV}^2\,, \quad
        m_\nu = 9.5179\times 10^{-3} \text{ eV}\,, \nonumber\\
    & \frac{M^\nu_D}{10^{-2}\text{~eV}} = \left(
    \begin{array}{ccc}
       1.56882+0.0048i & 0 & 0 \\
       0 & 2.17863+0.01456i & 0 \\
       -0.74382-0.05517i & 3.06548+0.02102i & 3.93054+0.02678i
    \end{array}
    \right)\,, \\
    & s^2_{12} = 0.3188\,, \quad  s^2_{23} = 0.4708\,, \quad  s^2_{13} = 0.02248\,, \quad
        \delta^{CP} = 180.0\degree\,, \quad J = 0.0 \,, \nonumber\\
    & \Delta m^2_{21} = 7.556 \times 10^{-5} \text{ eV}^2\,, \quad \Delta m^2_{31} = 2.521 \times 10^{-3} \text{ eV}^2\,, \quad
        m_\nu = 1.7371\times 10^{-2} \text{ eV}\,. \nonumber
\end{align}

\subsubsection{Comparison of $n$-zero Texture Searches}
\label{sec:nzerocomparison}

The two methods discussed in Sec.~\ref{sec:discoverzero} to search for $n$-zero textures
differ mainly in two ways, i.e., 
the selection of the initial parameters and the choice of loss function.
In the case described in Sec.~\ref{sec:nzerorandom},
we first randomly set parameter values in search of neutrino mass matrices where
$n$ elements have absolute values significantly smaller than the other $9 - n$.
Then, we use a loss function that directly minimizes only those $n$ elements.
For the method of Sec.~\ref{sec:nzerolosssearch},
we take the neutrino mass matrices as they come from the initialization
and apply a loss function that counts the number of elements below
and above some threshold.
Then, we apply a second minimization to make the $n$ smallest elements
as close to 0 as possible.
In both cases, we set conditions for the acceptable textures,
such as having no rows or columns full of vanishing elements
in the neutrino mass matrix.

The result of these discussions is that
we see that for the most common cases,i.e., 
1- and 2-zero textures,
the sizes of the elements are roughly the same
for both methods,
with the number of surviving matrices
having almost exactly the same proportions.
Furthermore, in the case of 3-zero textures,
we see that the search of Sec.~\ref{sec:nzerorandom}
yields mostly matrices with four values smaller than the rest.
This is not repeated in the case where we search for matrices
using optimization,
where the 3-zero texture cases all have
consistently three elements much smaller than the other six.
Additionally, in the top rows of Figs.~\ref{fig:commonrandom}
and~\ref{fig:commonlosssearch},
we see that for 2-zero textures the three most common configurations are actually the same
in both searches.
When comparing the three most common configurations for 3-zero textures,
shown in the second row of the same figures,
we see that the most common 3-zero texture found in Sec.~\ref{sec:nzerorandom}
matches the third most common 3-zero texture found in Sec.~\ref{sec:nzerolosssearch},
while the other two configurations are different.
Perhaps the most interesting difference
is that the search using optimization
actually yields usable 4-zero textures
at a rate of 10 out of 50,
while the random search yields no usable matrix
out of 500,
indicating that 4-zero textures
are not easy to find from random searches.
One problematic point of the search using optimization
is the inefficiency of performing first one optimization
to identify the number of vanishing elements
and then one more optimization to obtain a matrix
where those elements are significantly smaller than the others.

\section{Discussion}
\label{sec:discussion}

Up to this point, we have made a connection between ML techniques
and elements in a simple SM extension to obtain Dirac neutrino mass matrices.
While the purpose of this work is focused on methods and their applications
as a proof of concept and demonstration,
the processes detailed here can be adapted for further more specific discussions
concerning different origins of neutrino masses.
A simple example is already seen from the initialization itself,
where differences between elements of the mass matrix elements
already suggest a symmetry between $\mu-\tau$ families.
Moreover, there is an interesting tendency to have
similarly valued $M^\nu_{D12}$ and $M^\nu_{D13}$
which is not repeated for $M^\nu_{D21}$ and $M^\nu_{D31}$,
and happens more commonly than any other differences.
However, what may be the most interesting use of ML technique
is the design of loss functions
that could automatically use optimization to look for suitable cases.
One example of this is the search of $n$-zero textures by
counting (non-)vanishing mass matrix elements.
Finding specific forms of mass matrices could be used to inform
the search of symmetries.
One could start with a much more general setup,
allowing many couplings, mass terms, and parameters,
and then design a suitable loss function to find coincidences
or unnecessary parameters.
This could lead to a simpler model that is driven by current data.
Additionally
in our case, the parameters from the model have been included in the optimization.
Therefore, optimization has given us information on suitable choices of parameters,
as well as disfavored regions.  It has also revealed parameters that make no difference
and have no effect on the optimization process.
This could be employed in the aforementioned point
about deciding which parameters are really relevant for the model.

The methodology contained in this work can be a step towards more AI-inspired improvements.
Namely, given the type of data generated here, we can envision future extensions
where we could apply symbolic regression to infer more involved relationships
between parameters or between mass elements,
resulting in symbolic expressions that could inspire theoretical formulations,
or at least simplify searches for suitable parameters via other numerical methods.
Additionally, the parameters and particular forms of neutrino mass matrices
that were obtained here
can be used to train deep neural networks
that would be able to explore the parameter space much faster
and predict where certain forms of neutrino masses could be found.
This would certainly be useful in models with a large number of parameters
as it could learn the relationship between model parameters, predictions,
and desirable features,
allowing fast predictions of suitable parameters
that would later be checked against the full calculation.
We have left some of these opportunities
as possible future extensions of this work
which we plan to explore.

\section{Conclusion}
\label{sec:conclusion}

Neutrinos remain one of the most mysterious particles of the SM of particle physics.
They are part of it, but their description remains incomplete.
Numerous experiments indicate that neutrinos are massive
and oscillate while they propagate.  Additionally, future experiments promise to further constrain the involved
observables.
With no leading candidate to extend the SM to explain neutrino masses,
one may wonder if by developing new methods to relate available
results and proposed theories it could be possible to shed light on properties
that may not be found using traditional methods.
While we do not claim to have found any of these new properties,
we have presented a method for the applications of ML techniques
to the objects involved in BSM extensions for massive (Dirac) neutrinos.
We have explained the role played by tools
such as initialization, optimization, and the loss function,
as well as presented different loss functions
with the objective of exploring different results.
This permitted us to explore theoretical variables
and shapes of the neutrino mass matrix
that are allowed by constraints from neutrino data,
namely, from oscillation parameters and neutrino mass scale.
We started with a random baseline,
which practically defined our proposed initialization.
From there, we moved optimization of neutrino mass matrices
using specific loss functions crafted to achieve some textures
with different degrees of specificity.
In some cases, we specified exactly which elements we wanted to be equal
or wanted to vanish,
but in another cases we let the optimization itself decide
how many and which parameters should vanish based on experimental constraints.
We see this proposal as one step more
in opening possible ways of thinking about the relationship between
our observations and the theories we propose to explain them.
This is thanks to the development of new methods and technologies
aimed at letting computers explore different possibilities in automated ways,
guided by objectives that have been decided by researchers.
Furthermore, this can also be seen as one step towards methods
that may one day employ actual artificial intelligence
in the discovery of new and interesting relationships from data
as we obtain from experiments.

\section*{Acknowledgements}

AA and AS acknowledge support from SNII-SECIHTI (Mexico).
RR is supported by a KIAS Individual Grant (QP094602)
via the Quantum Universe Center
at Korea Institute for Advanced Study
and is deeply grateful to Facultad de Ciencias, Universidad de Colima
for its hospitality while performing part of this work.
Computational resources used in this work
have been supported by
the Center for Advanced Computation
at Korea Institute for Advanced Study.

\providecommand{\href}[2]{#2}\begingroup\raggedright\endgroup


\begin{thebibliography}{10}

\bibitem{ParticleDataGroup:2024cfk}
{\scshape Particle Data Group} collaboration, \emph{{Review of particle physics}}, \href{https://doi.org/10.1103/PhysRevD.110.030001}{\emph{Phys. Rev. D} {\bfseries 110} (2024) 030001}.

\bibitem{Mann:1976mp}
A.K.~Mann and H.~Primakoff, \emph{{Neutrino Oscillations and the Number of Neutrino Types}}, \href{https://doi.org/10.1103/PhysRevD.15.655}{\emph{Phys. Rev. D} {\bfseries 15} (1977) 655}.

\bibitem{Wolfenstein:1978uw}
L.~Wolfenstein, \emph{{Oscillations Among Three Neutrino Types and CP Violation}}, \href{https://doi.org/10.1103/PhysRevD.18.958}{\emph{Phys. Rev. D} {\bfseries 18} (1978) 958}.

\bibitem{Harrison:2002er}
P.F.~Harrison, D.H.~Perkins and W.G.~Scott, \emph{{Tri-bimaximal mixing and the neutrino oscillation data}}, \href{https://doi.org/10.1016/S0370-2693(02)01336-9}{\emph{Phys. Lett. B} {\bfseries 530} (2002) 167} [\href{https://arxiv.org/abs/hep-ph/0202074}{{\ttfamily hep-ph/0202074}}].

\bibitem{Harrison:2002kp}
P.F.~Harrison and W.G.~Scott, \emph{{Symmetries and generalizations of tri - bimaximal neutrino mixing}}, \href{https://doi.org/10.1016/S0370-2693(02)01753-7}{\emph{Phys. Lett. B} {\bfseries 535} (2002) 163} [\href{https://arxiv.org/abs/hep-ph/0203209}{{\ttfamily hep-ph/0203209}}].

\bibitem{Xing:2002sw}
Z.-z.~Xing, \emph{{Nearly tri bimaximal neutrino mixing and CP violation}}, \href{https://doi.org/10.1016/S0370-2693(02)01649-0}{\emph{Phys. Lett. B} {\bfseries 533} (2002) 85} [\href{https://arxiv.org/abs/hep-ph/0204049}{{\ttfamily hep-ph/0204049}}].

\bibitem{He:2003rm}
X.G.~He and A.~Zee, \emph{{Some simple mixing and mass matrices for neutrinos}}, \href{https://doi.org/10.1016/S0370-2693(03)00390-3}{\emph{Phys. Lett. B} {\bfseries 560} (2003) 87} [\href{https://arxiv.org/abs/hep-ph/0301092}{{\ttfamily hep-ph/0301092}}].

\bibitem{Xing:2006xa}
Z.-z.~Xing, H.~Zhang and S.~Zhou, \emph{{Nearly Tri-bimaximal Neutrino Mixing and CP Violation from mu-tau Symmetry Breaking}}, \href{https://doi.org/10.1016/j.physletb.2006.08.045}{\emph{Phys. Lett. B} {\bfseries 641} (2006) 189} [\href{https://arxiv.org/abs/hep-ph/0607091}{{\ttfamily hep-ph/0607091}}].

\bibitem{DayaBay:2012fng}
{\scshape Daya Bay} collaboration, \emph{{Observation of electron-antineutrino disappearance at Daya Bay}}, \href{https://doi.org/10.1103/PhysRevLett.108.171803}{\emph{Phys. Rev. Lett.} {\bfseries 108} (2012) 171803} [\href{https://arxiv.org/abs/1203.1669}{{\ttfamily 1203.1669}}].

\bibitem{RENO:2012mkc}
{\scshape RENO} collaboration, \emph{{Observation of Reactor Electron Antineutrino Disappearance in the RENO Experiment}}, \href{https://doi.org/10.1103/PhysRevLett.108.191802}{\emph{Phys. Rev. Lett.} {\bfseries 108} (2012) 191802} [\href{https://arxiv.org/abs/1204.0626}{{\ttfamily 1204.0626}}].

\bibitem{DoubleChooz:2014kuw}
{\scshape Double Chooz} collaboration, \emph{{Improved measurements of the neutrino mixing angle $\theta_{13}$ with the Double Chooz detector}}, \href{https://doi.org/10.1007/JHEP02(2015)074}{\emph{JHEP} {\bfseries 10} (2014) 086} [\href{https://arxiv.org/abs/1406.7763}{{\ttfamily 1406.7763}}].

\bibitem{Matchev:2023mii}
K.T.~Matchev, K.~Matcheva, P.~Ramond and S.~Verner, \emph{{Seeking Truth and Beauty in Flavor Physics with Machine Learning}},  in \emph{{37th Conference on Neural Information Processing Systems}}, 10, 2023 [\href{https://arxiv.org/abs/2311.00087}{{\ttfamily 2311.00087}}].

\bibitem{Matchev:2024ash}
K.T.~Matchev, K.~Matcheva, P.~Ramond and S.~Verner, \emph{{Exploring the truth and beauty of theory landscapes with machine learning}}, \href{https://doi.org/10.1016/j.physletb.2024.138941}{\emph{Phys. Lett. B} {\bfseries 856} (2024) 138941} [\href{https://arxiv.org/abs/2401.11513}{{\ttfamily 2401.11513}}].

\bibitem{Nishimura:2020nre}
S.~Nishimura, C.~Miyao and H.~Otsuka, \emph{{Exploring the flavor structure of quarks and leptons with reinforcement learning}}, \href{https://doi.org/10.1007/JHEP12(2023)021}{\emph{JHEP} {\bfseries 23} (2020) 021} [\href{https://arxiv.org/abs/2304.14176}{{\ttfamily 2304.14176}}].

\bibitem{Nishimura:2024apb}
S.~Nishimura, C.~Miyao and H.~Otsuka, \emph{{Reinforcement learning-based statistical search strategy for an axion model from flavor}}, \href{https://doi.org/10.1007/JHEP10(2025)043}{\emph{JHEP} {\bfseries 10} (2025) 043} [\href{https://arxiv.org/abs/2409.10023}{{\ttfamily 2409.10023}}].

\bibitem{Nishimura:2025rsk}
S.~Nishimura, H.~Otsuka and H.~Uchiyama, \emph{{Exploring the flavor structure of leptons via diffusion models}}, \href{https://doi.org/10.1103/rtnd-vwt9}{\emph{Phys. Rev. D} {\bfseries 113} (2026) 055030} [\href{https://arxiv.org/abs/2503.21432}{{\ttfamily 2503.21432}}].

\bibitem{Nishimura:2025knz}
S.~Nishimura, H.~Otsuka and H.~Uchiyama, \emph{{Diffusion-Model Approach to Flavor Models: A Case Study for $S'_4$ Modular Flavor Model}}, \href{https://doi.org/10.1093/ptep/ptag069}{\emph{PTEP} {\bfseries 2026} (2026) 053B08} [\href{https://arxiv.org/abs/2504.00944}{{\ttfamily 2504.00944}}].

\bibitem{Giarnetti:2025mit}
A.~Giarnetti and D.~Meloni, \emph{{Reinforcement Learning Techniques for the Flavor Problem in Particle Physics}}, \href{https://doi.org/10.3390/sym18010131}{\emph{Symmetry} {\bfseries 18} (2026) 131} [\href{https://arxiv.org/abs/2510.25495}{{\ttfamily 2510.25495}}].

\bibitem{ParticleDataGroup:2018ovx}
{\scshape Particle Data Group} collaboration, \emph{{Review of Particle Physics}}, \href{https://doi.org/10.1103/PhysRevD.98.030001}{\emph{Phys. Rev. D} {\bfseries 98} (2018) 030001}.

\bibitem{Jarlskog:1985ht}
C.~Jarlskog, \emph{{Commutator of the Quark Mass Matrices in the Standard Electroweak Model and a Measure of Maximal CP Nonconservation}}, \href{https://doi.org/10.1103/PhysRevLett.55.1039}{\emph{Phys. Rev. Lett.} {\bfseries 55} (1985) 1039}.

\bibitem{Wu:1985ea}
D.-d.~Wu, \emph{{The Rephasing Invariants and CP}}, \href{https://doi.org/10.1103/PhysRevD.33.860}{\emph{Phys. Rev. D} {\bfseries 33} (1986) 860}.

\bibitem{KATRIN:2024cdt}
{\scshape KATRIN} collaboration, \emph{{Direct neutrino-mass measurement based on 259 days of KATRIN data}}, \href{https://doi.org/10.1126/science.adq9592}{\emph{Science} {\bfseries 388} (2025) adq9592} [\href{https://arxiv.org/abs/2406.13516}{{\ttfamily 2406.13516}}].

\bibitem{Esteban:2024eli}
I.~Esteban, M.C.~Gonzalez-Garcia, M.~Maltoni, I.~Martinez-Soler, J.P.~Pinheiro and T.~Schwetz, \emph{{NuFit-6.0: updated global analysis of three-flavor neutrino oscillations}}, \href{https://doi.org/10.1007/JHEP12(2024)216}{\emph{JHEP} {\bfseries 12} (2024) 216} [\href{https://arxiv.org/abs/2410.05380}{{\ttfamily 2410.05380}}].

\bibitem{nufitwebsite}
{\scshape NuFIT} collaboration, ``{NuFIT v6.1 (2025)}.'' \url{http://www.nu-fit.org/}, 2025.

\bibitem{Fukuyama:1997ky}
T.~Fukuyama and H.~Nishiura, \emph{{Mass matrix of Majorana neutrinos}},  \href{https://arxiv.org/abs/hep-ph/9702253}{{\ttfamily hep-ph/9702253}}.

\bibitem{Ma:2001mr}
E.~Ma and M.~Raidal, \emph{{Neutrino mass, muon anomalous magnetic moment, and lepton flavor nonconservation}}, \href{https://doi.org/10.1103/PhysRevLett.87.011802}{\emph{Phys. Rev. Lett.} {\bfseries 87} (2001) 011802} [\href{https://arxiv.org/abs/hep-ph/0102255}{{\ttfamily hep-ph/0102255}}].

\bibitem{Balaji:2001ex}
K.R.S.~Balaji, W.~Grimus and T.~Schwetz, \emph{{The Solar LMA neutrino oscillation solution in the Zee model}}, \href{https://doi.org/10.1016/S0370-2693(01)00532-9}{\emph{Phys. Lett. B} {\bfseries 508} (2001) 301} [\href{https://arxiv.org/abs/hep-ph/0104035}{{\ttfamily hep-ph/0104035}}].

\bibitem{Lam:2001fb}
C.S.~Lam, \emph{{A 2-3 symmetry in neutrino oscillations}}, \href{https://doi.org/10.1016/S0370-2693(01)00465-8}{\emph{Phys. Lett. B} {\bfseries 507} (2001) 214} [\href{https://arxiv.org/abs/hep-ph/0104116}{{\ttfamily hep-ph/0104116}}].

\bibitem{Harrison:2002et}
P.F.~Harrison and W.G.~Scott, \emph{{$\mu$-$\tau$ reflection symmetry in lepton mixing and neutrino oscillations}}, \href{https://doi.org/10.1016/S0370-2693(02)02772-7}{\emph{Phys. Lett. B} {\bfseries 547} (2002) 219} [\href{https://arxiv.org/abs/hep-ph/0210197}{{\ttfamily hep-ph/0210197}}].

\bibitem{Xing:2015fdg}
Z.-z.~Xing and Z.-h.~Zhao, \emph{{A review of {\ensuremath{\mu}}-{\ensuremath{\tau}} flavor symmetry in neutrino physics}}, \href{https://doi.org/10.1088/0034-4885/79/7/076201}{\emph{Rept. Prog. Phys.} {\bfseries 79} (2016) 076201} [\href{https://arxiv.org/abs/1512.04207}{{\ttfamily 1512.04207}}].

\bibitem{kingma2017adammethodstochasticoptimization}
D.P.~Kingma and J.~Ba, \emph{Adam: A method for stochastic optimization},  \href{https://arxiv.org/abs/1412.6980}{{\ttfamily 1412.6980}}.

\bibitem{Fritzsch:2011qv}
H.~Fritzsch, Z.-z.~Xing and S.~Zhou, \emph{{Two-zero Textures of the Majorana Neutrino Mass Matrix and Current Experimental Tests}}, \href{https://doi.org/10.1007/JHEP09(2011)083}{\emph{JHEP} {\bfseries 09} (2011) 083} [\href{https://arxiv.org/abs/1108.4534}{{\ttfamily 1108.4534}}].

\bibitem{Meloni:2014yea}
D.~Meloni, A.~Meroni and E.~Peinado, \emph{{Two-zero Majorana textures in the light of the Planck results}}, \href{https://doi.org/10.1103/PhysRevD.89.053009}{\emph{Phys. Rev. D} {\bfseries 89} (2014) 053009} [\href{https://arxiv.org/abs/1401.3207}{{\ttfamily 1401.3207}}].

\bibitem{Nguyen:2014mwa}
T.P.~Nguyen, \emph{{Texture zeros of neutrino mass matrix with seesaw mechanism and leptogenesis}}, \href{https://doi.org/10.1142/S0217732314500382}{\emph{Mod. Phys. Lett. A} {\bfseries 29} (2014) 1450038}.

\bibitem{Benavides:2022hca}
R.H.~Benavides, D.V.~Forero, L.~Mu{\~n}oz, J.M.~Mu{\~n}oz, A.~Rico and A.~Tapia, \emph{{Five texture zeros in the lepton sector and neutrino oscillations at DUNE}}, \href{https://doi.org/10.1103/PhysRevD.107.036008}{\emph{Phys. Rev. D} {\bfseries 107} (2023) 036008} [\href{https://arxiv.org/abs/2207.04072}{{\ttfamily 2207.04072}}].

\bibitem{Ding:2022aoe}
G.-J.~Ding, F.R.~Joaquim and J.-N.~Lu, \emph{{Texture-zero patterns of lepton mass matrices from modular symmetry}}, \href{https://doi.org/10.1007/JHEP03(2023)141}{\emph{JHEP} {\bfseries 03} (2023) 141} [\href{https://arxiv.org/abs/2211.08136}{{\ttfamily 2211.08136}}].

\bibitem{Devi:2023vpe}
M.R.~Devi, \emph{{Retrieving texture zeros in 3+1 active-sterile neutrino framework under the action of $A_4$ modular-invariants}},  \href{https://arxiv.org/abs/2303.04900}{{\ttfamily 2303.04900}}.

\bibitem{Rico:2023drr}
A.~Rico, R.H.~Benavides, D.V.~Forero, L.~Mu{\~n}oz and A.~Tapia, \emph{{Five texture zeros in the lepton sector that reproduce the observable neutrino masses and mixings: A mapping between the standard parameterization and our parameterization}}, \href{https://doi.org/10.22323/1.444.1047}{\emph{PoS} {\bfseries ICRC2023} (2023) 1047}.

\bibitem{Zhao:2023vkb}
Z.-h.~Zhao, \emph{{Combinations of the $\mu$-$\tau$ reflection symmetry and texture zeros in the Dirac neutrino mass matrix of the seesaw model}}, \href{https://doi.org/10.1140/epjp/s13360-023-04678-8}{\emph{Eur. Phys. J. Plus} {\bfseries 138} (2023) 1055}.

\bibitem{Kumar:2023iaj}
S.~Kumar and R.R.~Gautam, \emph{{Neutrino mass matrices with generalized CP symmetries and texture zeros}}, \href{https://doi.org/10.1016/j.nuclphysb.2024.116520}{\emph{Nucl. Phys. B} {\bfseries 1001} (2024) 116520} [\href{https://arxiv.org/abs/2312.07150}{{\ttfamily 2312.07150}}].

\bibitem{Zhou:2024gcp}
S.~Zhou, \emph{{Texture zeros for lepton flavor mixing}}, \href{https://doi.org/10.1142/S0217751X24410148}{\emph{Int. J. Mod. Phys. A} {\bfseries 39} (2024) 2441014}.

\bibitem{Nomura:2023usj}
T.~Nomura, H.~Okada and H.~Otsuka, \emph{{Texture zeros realization in a three-loop radiative neutrino mass model from modular A4 symmetry}}, \href{https://doi.org/10.1016/j.nuclphysb.2024.116579}{\emph{Nucl. Phys. B} {\bfseries 1004} (2024) 116579} [\href{https://arxiv.org/abs/2309.13921}{{\ttfamily 2309.13921}}].

\bibitem{Carrillo-Monteverde:2024ntq}
A.~Carrillo-Monteverde, S.~G{\'o}mez-{\'A}vila and L.~L{\'o}pez-Lozano, \emph{{2-zeros texture and the Universal Texture Constraint in the Leptonic Sector}}, \href{https://doi.org/10.1142/S0217751X24500520}{\emph{Int. J. Mod. Phys. A} {\bfseries 39} (2024) 2450052}.

\bibitem{Singh:2024ivf}
L.~Singh, M.~Kashav and S.~Verma, \emph{{Implications of~Dark-$\theta _{12}$ Solution on Two-zero Texture Inverse Neutrino Mass Matrix}}, \href{https://doi.org/10.1007/978-981-97-0289-3_157}{\emph{Springer Proc. Phys.} {\bfseries 304} (2024) 675}.

\bibitem{Mazumder:2024lne}
I.A.~Mazumder and R.~Dutta, \emph{{Bottom-up approach to texture zeros in the neutrino mass matrix}}, \href{https://doi.org/10.1142/S0217751X26500107}{\emph{Int. J. Mod. Phys. A} {\bfseries 41} (2026) 2650010} [\href{https://arxiv.org/abs/2409.04756}{{\ttfamily 2409.04756}}].

\bibitem{Kobayashi:2024cvp}
T.~Kobayashi, H.~Otsuka and M.~Tanimoto, \emph{{Yukawa textures from non-invertible symmetries}}, \href{https://doi.org/10.1007/JHEP12(2024)117}{\emph{JHEP} {\bfseries 12} (2024) 117} [\href{https://arxiv.org/abs/2409.05270}{{\ttfamily 2409.05270}}].

\bibitem{Kobayashi:2025znw}
T.~Kobayashi, Y.~Nishioka, H.~Otsuka and M.~Tanimoto, \emph{{More about quark Yukawa textures from selection rules without group actions}}, \href{https://doi.org/10.1007/JHEP05(2025)177}{\emph{JHEP} {\bfseries 05} (2025) 177} [\href{https://arxiv.org/abs/2503.09966}{{\ttfamily 2503.09966}}].

\bibitem{Kobayashi:2025ldi}
T.~Kobayashi, H.~Otsuka, M.~Tanimoto and H.~Uchida, \emph{{Lepton mass textures from non-invertible multiplication rules}}, \href{https://doi.org/10.1007/JHEP08(2025)189}{\emph{JHEP} {\bfseries 08} (2025) 189} [\href{https://arxiv.org/abs/2505.07262}{{\ttfamily 2505.07262}}].

\bibitem{Calibbi:2025ded}
L.~Calibbi, X.~Gao and M.~Yuan, \emph{{Hunting for neutrino texture zeros with muon and tau flavor violation}}, \href{https://doi.org/10.1007/JHEP05(2026)188}{\emph{JHEP} {\bfseries 05} (2026) 188} [\href{https://arxiv.org/abs/2511.08679}{{\ttfamily 2511.08679}}].

\bibitem{Biswas:2025woq}
A.~Biswas, S.~Jangid and S.C.~Park, \emph{{Investigating two-zero texture in the light of gauged Type-II seesaw}}, \href{https://doi.org/10.1016/j.physletb.2025.139786}{\emph{Phys. Lett. B} {\bfseries 868} (2025) 139786} [\href{https://arxiv.org/abs/2504.10312}{{\ttfamily 2504.10312}}].

\bibitem{Gogoi:2025htb}
J.~Gogoi and M.K.~Das, \emph{{A study of texture zeros using A4 discrete symmetry group}}, \href{https://doi.org/10.1016/j.jspc.2025.100071}{\emph{J. Subatomic Part. Cosmol.} {\bfseries 3} (2025) 100071}.

\bibitem{Jiang:2025psz}
Z.~Jiang, B.-Y.~Qu and G.-J.~Ding, \emph{{Texture-zeros in minimal seesaw from noninvertible symmetry fusion rules}}, \href{https://doi.org/10.1103/d29s-cw34}{\emph{Phys. Rev. D} {\bfseries 112} (2025) 115029} [\href{https://arxiv.org/abs/2510.07236}{{\ttfamily 2510.07236}}].

\bibitem{Borah:2025vtn}
D.~Borah, P.~Das and D.~Dutta, \emph{{Neutrino texture-zeros after JUNO's first results: Implications for long-baseline neutrino experiments}},  \href{https://arxiv.org/abs/2512.13587}{{\ttfamily 2512.13587}}.

\bibitem{Priya:2026tpk}
Priya, R.~Kumar, L.~Singh and S.~Verma, \emph{{Implications of the First JUNO Results for Dirac Neutrino Texture Zeros}},  \href{https://arxiv.org/abs/2604.19122}{{\ttfamily 2604.19122}}.

\end{thebibliography}
\end{document}